\documentclass[notitlepage,nofootinbib,superscriptaddress,preprintnumbers]{revtex4-1}
\pdfoutput=1
\usepackage[latin9]{inputenc}
\usepackage{amsmath}
\usepackage{amssymb}
\usepackage{graphicx}
\usepackage{epstopdf}
\usepackage{color}
\usepackage[hidelinks]{hyperref}

\makeatletter
\def\Mpl{M_{\rm P}}

\makeatother

\begin{document}

\preprint{YITP-16-140, IPMU16-0204}

\title{Minimal theory of quasidilaton massive gravity}

\author{Antonio De Felice}
\affiliation{Center for Gravitational Physics, Yukawa Institute for Theoretical Physics, Kyoto University, 606-8502, Kyoto, Japan}
\author{Shinji Mukohyama}
\affiliation{Center for Gravitational Physics, Yukawa Institute for Theoretical Physics, Kyoto University, 606-8502, Kyoto, Japan}
\affiliation{Kavli Institute for the Physics and Mathematics of the Universe (WPI), The University of Tokyo Institutes for Advanced Study, The University of Tokyo, Kashiwa, Chiba 277-8583, Japan}
\author{Michele Oliosi}
\affiliation{Center for Gravitational Physics, Yukawa Institute for Theoretical Physics, Kyoto University, 606-8502, Kyoto, Japan}

\date{\today}

 \begin{abstract}
We introduce a quasidilaton scalar field to the minimal theory of massive gravity with the Minkowski fiducial metric, in such a way that the quasidilaton global symmetry is maintained and that the theory admits a stable self-accelerating de Sitter solution. We start with a precursor theory that contains three propagating gravitational degrees of freedom without a quasidilaton scalar and introduce St\"{u}ckelberg fields to covariantize its action. This makes it possible for us to formulate the quasidilaton global symmetry that mixes the St\"{u}ckelberg fields and the quasidilaton scalar field. By the Hamiltonian analysis we confirm that the precursor theory with the quasidilaton scalar contains 4 degrees of freedom, three from the precursor massive gravity and one from the quasidilaton scalar. We further remove one propagating degree of freedom to construct the minimal quasidilaton theory with three propagating degrees of freedom, corresponding to two polarizations of gravitational waves from the minimal theory of massive gravity and one scalar from the quasidilaton field, by carefully introducing two additional constraints to the system in the Hamiltonian language. Switching to the Lagrangian language, we find self-accelerating de Sitter solutions in the minimal quasidilaton theory and analyze their stability. It is found that the self-accelerating de Sitter solution is stable in a wide range of parameters. 
 \end{abstract}

\maketitle

\section{Introduction}
General relativity has been lately more successful than ever, see for instance the recent direct observation of gravitational waves \cite{bib:LIGO} by the LIGO Collaboration. However, we still cryingly lack a satisfactory explanation for numerous holes in our understanding of the Universe. One of these holes is the accelerated expansion of the Universe which, in the standard lore, can be accounted by a bare cosmological constant. This current view is probably just a temporary placeholder for a more fundamental explanation; it is for instance still unclear how and whether vacuum energy is to be taken into account, considering the relatively small value of the acceleration or cosmological constant \cite{Burgess:2013ara}.

The accelerated expansion is a large distance phenomenon. A direct way to tackle the problem is thus to depart from general relativity in the IR, for instance, by adding a small mass to the graviton. Were the resulting models to conclude in favor of an expendable bare cosmological constant, or a screening of a large \textemdash natural \textemdash cosmological constant, we would indeed have found an appreciable solution to the puzzle \cite{Tolley:2015oxa}. More generally, the possibility of the graviton having a mass, or any other modification in the IR, changes the rules of late-time cosmology, thus potentially offering new paths to a better understanding of late-time accelerated expansion. 

In addition to the phenomenological applications of considering alternatives to general relativity, the quest for a viable theory of massive gravity is also of theoretical interest, as it has proved since its early years to be a challenging task. Fierz and Pauli were the first, in 1939, to propose a theory for a massive spin-2 field on a Minkowski background \cite{bib:Pauli-Fierz}. Many decades later, theories of massive gravity were shown however not to coincide with general relativity in their massless limit \cite{bib:vdvz}, an issue known as the vDVZ discontinuity, and to be accompanied by a ghost \cite{Boulware:1973my}, the Boulware-Deser ghost. Although the first problem was quickly recognized by Vainshtein to be an artifact of the linearity of the theory \cite{Vainshtein:1972sx}, the second had to wait until the ghost-free massive gravity model \cite{deRham:2010kj} by de Rham, Gabadadze, and Tolley (dRGT), in 2010, to be cured. It was then quickly shown that viable cosmology remained difficult \textemdash even in the at the time newly found dRGT model \cite{DeFelice:2012mx}. Several routes have been since explored such as breaking either homogeneity or isotropy of the solution at the background level \cite{DAmico:2011eto,Gumrukcuoglu:2012aa} or further modifying the original theory \cite{DAmico:2012hia, Huang:2012pe, DeFelice:2013tsa, deRham:2014gla, DeFelice:2015yha}.

In this paper we combine two ideas of modification of the original dRGT model. The first of these two ideas was introduced in an intent to add a scalar field to the dRGT theory \cite{DAmico:2012hia}. The form of the action is restricted by a scaling-type global symmetry and the additional scalar field resembles a dilaton scalar field to some extent. For this reason the added scalar field is often called the quasidilaton. 
Although it was eventually recognized that the self-accelerating
cosmological solutions were in general unstable in the original
quasidilaton theory \cite{Gumrukcuoglu:2013nza,D'Amico:2013kya}, further extensions have been developed \cite{DeFelice:2013dua, DeFelice:2013tsa, Mukohyama:2013raa,Mukohyama:2014rca,DeFelice:2016tiu,Gumrukcuoglu:2016hic,Kahniashvili:2014wua,Motohashi:2014una,Heisenberg:2015voa,Gumrukcuoglu:2017ioy}. (See also \cite{Gabadadze:2014kaa} for a possible new type of
solutions in the decoupling limit.) In particular it was recently
shown that there exists a ghost-free quasidilaton massive gravity
allowing for a stable self-accelerating cosmological solution \cite{Gumrukcuoglu:2017ioy}.
 The second of the two ideas is to remove the problematic degrees of freedom by imposing adequate constraints and to obtain a theory of massive gravity with, as in general relativity, two tensor modes only \cite{DeFelice:2015hla, DeFelice:2015moy,DeFelice:2016ufg}. The theory constructed in this way is thus called the minimal theory of massive gravity (MTMG) and provides a nonlinear completion of the self-accelerating solution~\cite{Gumrukcuoglu:2011ew,Gumrukcuoglu:2011zh} found in the original dRGT theory. 

By combining the aforementioned two approaches we obtain a theory with three degrees of freedom, one scalar due to the quasidilaton extension, supplemented by the remaining two tensor modes of the MTMG. We thus call this theory the \textit{minimal quasidilaton}. The theory of minimal quasidilaton that we shall develop in the present paper allows for a self-accelerating de Sitter solution with stable dynamics of perturbations, in a wide region of the parameter space. We illustrate this by presenting in detail the allowed parameter space in some chosen cases.

As a feature, our theory inherits the Lorentz violation of the MTMG. This is a necessary requirement for reducing the number of degrees of freedom with respect to a Lorentz invariant theory of a massive spin-2 field and thus eliminating unwanted helicity-$0$ and helicity-$1$ degrees. From a phenomenological point of view, this is acceptable as long as the violation is small enough to satisfy various constraints~\cite{Kostelecky:2008ts}. Our model indeed satisfies this requirement, as the violation appears only in the gravity sector at the scale $1/m$ or longer, where $m$ is the graviton mass and is set to be of order the present Hubble expansion rate. From this viewpoint, the model belongs to Lorentz-violating massive gravity theories. However, as in the MTMG, and contrary to the theories studied previously in \cite{ArkaniHamed:2003uy, Rubakov:2004eb, Dubovsky:2004sg, Berezhiani:2007zf, Blas:2007ep, Grisa:2008um, Blas:2009my, Comelli:2014xga}, not only the potential structure but also the kinetic part of the Lagrangian is changed and breaks Lorentz invariance at the cosmological scale. In the case of the MTMG, this is the reason why the theory provides a stable nonlinear completion of the self-accelerating cosmological solution that was originally found in the dRGT theory. For the same reason, our theory studied in the present paper can accommodate a stable scaling-type cosmological solution that also self-accelerates the expansion of the Universe. Finally from the point of view of a possible UV completion of the theory, it has been known that Lorentz invariance can be broken either spontaneously or explicitly in quantum gravity candidates such as superstring theory~\cite{Kostelecky:1988zi,Douglas:2001ba}, loop quantum gravity~\cite{Gambini:1998it,Alfaro:2001rb} and Ho\v{r}ava-Lifshitz gravity~\cite{Horava:2009uw,Mukohyama:2010xz} (see also \cite{Janiszewski:2012nf}). Lorentz violation can also appear as a low energy effective feature of Lorentz invariant theories by a spontaneous symmetry breaking.

While the MTMG \cite{DeFelice:2015hla}, the extended quasidilaton \cite{DeFelice:2013dua} and the new quasidilaton \cite{Mukohyama:2014rca} themselves provide stable cosmological solutions with self-acceleration, the minimal quasidilaton theory possesses its own advantages. From the viewpoint of extending the quasidilaton theory, the minimal quasidilaton theory has a smaller number of propagating degrees of freedom and thus the stability is easier to establish compared to other extensions. From the viewpoint of extending the MTMG, the choice of the fiducial metric/vielbein in the minimal quasidilaton theory is simpler than the original theory. Extending on this argument, the minimal quasidilaton theory can be seen as a first step towards a theory of minimal bigravity theory, where the fiducial metric is not anymore chosen as a definition of the theory but is a full-fledged dynamical entity. Indeed, in the minimal quasidilaton model, the fiducial metric can be considered as partially dynamical via the quasidilaton scalar field. These advantages make it worthwhile investigating the theory of minimal quasidilaton in detail. 

The rest of the paper is organized as follows. In Sec.\ \ref{sec:precursor}, we start by presenting a short review of the original precursor theory that was introduced in \cite{DeFelice:2015hla} to construct the MTMG, in its unitary-gauge formulation. We then proceed to establish its covariant formulation via the introduction of St\"{u}ckelberg fields. The extension with the quasidilaton scalar field and the associated global symmetry is presented in Sec.\ \ref{sec:quasidilaton-precursor}. We next recover the unitary-gauge formulation of the new precursor theory with a quasidilaton field, by fixing the St\"{u}ckelberg fields. A Hamiltonian analysis is then performed, which allows us, in Sec.\ \ref{sec:hamiltonian}, to promote the precursor theory to the minimal theory by adding two constraints to the precursor Hamiltonian. In Sec.\ \ref{sec:lagrangian}, we return to the the Lagrangian picture via a Legendre transformation, and write down the action for the minimal quasidilaton theory both in the vielbein and in the metric formalisms. We then finally explore the behavior of the theory in de Sitter backgrounds (Sec.\ \ref{sec:desitter}) and their Minkowski limit (Sec.\ \ref{sec:minkowski}). The action for the minimal quasidilaton theory, as well as the analysis of background and linear perturbations in the self-accelerating de Sitter solution and its Minkowski limit, are the principal results of the paper.

\section{Precursor theory}\label{sec:precursor}

\subsection{Action in unitary gauge}

We review here the construction of the precursor theory for the MTMG in the unitary gauge, presented in \cite{DeFelice:2015moy}. We refer the reader to this reference for more details. 

The theory uses as basic ingredients the lapse function $N$, the shift vector $N^{i}$, the spatial vielbein $e^{I}{}_{j}$, and the corresponding fiducial quantities $M$, $M^{i}$, and $E^{I}{}_{j}$ ($i,j,\cdots = 1,2,3$ and $I,\cdots=1,2,3$). One also defines the dual basis $e_{I}{}^{j}$ and $E_{I}{}^{i}$, respectively, of both sets of vielbeins so that
\begin{equation}
e_{I}{}^{k}e^{J}{}_{k}=\delta_{I}^{J}\,,\quad e_{I}{}^{i}e^{I}{}_{j}=\delta_{j}^{i}\,,\quad E_{I}{}^{k}E^{J}{}_{k}=\delta_{I}^{J}\,,\quad E_{I}{}^{i}E^{I}{}_{j}=\delta_{j}^{i}\,.\label{eqn:def-einverse-unitary}
\end{equation}
In addition to these quantities, it is useful to introduce the two spatial metrics -- physical and fiducial -- and the combinations $\gamma_{ij}$, $\tilde{\gamma}_{ij}$, $X_{I}{}^{J}$ and $Y_{I}{}^{J}$, defined by
\begin{equation}
\gamma_{ij}=\delta_{IJ}e^{I}{}_{i}e^{J}{}_{j}\,,\quad\tilde{\gamma}_{ij}=\delta_{IJ}E^{I}{}_{i}E^{J}{}_{j}\,,
\end{equation}
and 
\begin{equation}
X_{I}{}^{J}\equiv e_{I}{}^{i}E^{J}{}_{i}\,,\quad Y_{I}{}^{J}\equiv E_{I}{}^{i}e^{J}{}_{i}\,.
\end{equation}
We denote the inverses of $\gamma_{ij}$ and $\tilde{\gamma}_{ij}$ as $\gamma^{ij}$ and $\tilde{\gamma}^{ij}$, respectively. Note that $X_{I}{}^{J}$ and $Y_{I}{}^{J}$ are constructed to be the inverse of each other, i.e.
\begin{equation}
X_{I}{}^{J} Y_{J}{}^{L} = \delta^{L}_{I}\,, \quad Y_{I}{}^{J} X_{J}{}^{L} = \delta^{L}_{I}\,.
\end{equation}
With these elements the precursor action is written as 
\begin{eqnarray}
S_{\mathrm{pre}} & = & \frac{M_{\mathrm{P}}^{2}}{2}\int d^{4}x \Bigl\{N \sqrt{\gamma}\left(R[\gamma_{ij}]+K_{ij}K^{ij}-K^2\right)\nonumber \\
&  & -c_{0}m^{2}\sqrt{\tilde{\gamma}}M-c_{1}m^{2}\sqrt{\tilde{\gamma}}(N+M Y_{I}{}^{I})\nonumber \\
 &  & -c_{2}m^{2}\sqrt{\tilde{\gamma}}\left[NY_{I}{}^{I}+\frac{M}{2}\,(Y_{I}{}^{I}Y_{J}{}^{J}-Y_{I}{}^{J}Y_{J}{}^{I})\right]\nonumber \\
 &  & -c_{3}m^{2}\sqrt{\gamma}(NX_{I}{}^{I}+M)-c_{4}m^{2}N\sqrt{\gamma}\Bigr\}\,,\label{eqn:action-precursor-unitary}
\end{eqnarray}
where $\sqrt{\gamma}=\sqrt{\det\gamma_{ij}}$, $\sqrt{\tilde{\gamma}}=\sqrt{\det\tilde{\gamma}_{ij}}$, $K_{ij}$ denotes the extrinsic curvature and $K=\gamma^{ij}K_{ij}$ is its trace, as common in the Arnowitt Deser Misner (ADM) formalism.

\subsection{Covariant precursor Lagrangian}

The basic variables of the precursor theory in its covariant formulation are the $4$-dimensional physical metric $g_{\mu\nu}$ and
the four scalars ($\phi^{0}$, $\phi^{1}$, $\phi^{2}$, $\phi^{3}$).
Out of these, we construct spacetime scalars $\boldsymbol{N}$, $\boldsymbol{N}^{p}$ and $\boldsymbol{\gamma}^{pq}$
($p,q,\cdots=1,2,3$) as 
\begin{equation}
\boldsymbol{N}=\frac{1}{\sqrt{-g^{\mu\nu}\partial_{\mu}\phi^{0}\partial_{\nu}\phi^{0}}}\,,\quad \boldsymbol{N}^{p}=\boldsymbol{N}^{2}g^{\mu\nu}\partial_{\mu}\phi^{0}\partial_{\nu}\phi^{p}\,,\quad\boldsymbol{\gamma}^{pq}=g^{\mu\nu}\partial_{\mu}\phi^{p}\partial_{\nu}\phi^{q}+\frac{\boldsymbol{N}^{p}\boldsymbol{N}^{q}}{\boldsymbol{N}^{2}}\,,\label{eqn:def-N-Np-gammapq}
\end{equation}
so that 
\begin{equation}
g^{\mu\nu}\partial_{\mu}\phi^{0}\partial_{\nu}\phi^{0}=-\frac{1}{\boldsymbol{N}^{2}}\,,\quad g^{\mu\nu}\partial_{\mu}\phi^{0}\partial_{\nu}\phi^{p}=\frac{\boldsymbol{N}^{p}}{\boldsymbol{N}^{2}}\,,\quad g^{\mu\nu}\partial_{\mu}\phi^{p}\partial_{\nu}\phi^{q}=\boldsymbol{\gamma}^{pq}-\frac{\boldsymbol{N}^{p}\boldsymbol{N}^{q}}{\boldsymbol{N}^{2}}\,.
\end{equation}
We also define a set of spacetime scalars $\boldsymbol{e}_{I}{}^{p}$ ($I,\cdots=1,2,3$)
satisfying 
\begin{equation}
\boldsymbol{\gamma}^{pq}=\delta^{IJ}\boldsymbol{e}_{I}{}^{p}\boldsymbol{e}_{J}{}^{q}\,.\label{eqn:def-eIp}
\end{equation}
This uniquely defines $\boldsymbol{e}_{I}{}^{p}$ up to an arbitrary orthogonal
transformation 
\begin{equation}
\boldsymbol{e}_{I}{}^{p}\to O_{I}{}^{J}\boldsymbol{e}_{J}{}^{p}\,,\quad\delta^{IJ}O_{I}{}^{K}O_{J}{}^{L}=\delta^{KL}\,.\label{eqn:orthogonal-tr}
\end{equation}
We then construct $\boldsymbol{\gamma}_{pq}$ and $\boldsymbol{e}^{I}{}_{p}$ as sets of spacetime scalars that form the inverse matrices of $\boldsymbol{\gamma}^{pq}$ and $\boldsymbol{e}_{I}{}^{p}$, respectively, as 
\begin{equation}
\boldsymbol{\gamma}^{pr}\boldsymbol{\gamma}_{rq}=\delta_{q}^{p}\,,\quad\boldsymbol{\gamma}_{pq}=\boldsymbol{\gamma}_{qp}\,,\label{eqn:def-gammainverse}
\end{equation}
and 
\begin{equation}
\boldsymbol{e}_{I}{}^{p}\boldsymbol{e}^{J}{}_{p}=\delta_{I}^{J}\,,\quad \boldsymbol{e}_{I}{}^{p}\boldsymbol{e}^{I}{}_{q}=\delta_{q}^{p}\,.\label{eqn:def-einverse}
\end{equation}

The theory also contains a fixed $\phi^{0}$-dependent function $\boldsymbol{M}(\phi^{0},\phi^{r})$
and a fixed $\phi^{0}$-dependent vielbein $\boldsymbol{E}^{I}{}_{p}(\phi^{0},\phi^{r})$
($I,\cdots=1,2,3$) in the $3$-dimensional field space spanned by
$\{\phi^{1},\phi^{2},\phi^{3}\}$. Out of them, we can construct a
fixed $\phi^{0}$-dependent metric $\tilde{\boldsymbol{\gamma}}_{pq}(\phi^{0},\phi^{r})$
in the $3$-dimensional field space as 
\begin{equation}
\tilde{\boldsymbol{\gamma}}_{pq}\equiv\delta_{IJ}\boldsymbol{E}^{I}{}_{p}\boldsymbol{E}^{J}{}_{q}\,.
\end{equation}
It is convenient to introduce the inverse metric $\tilde{\boldsymbol{\gamma}}^{pq}$
and the dual basis $\boldsymbol{E}_{I}{}^{p}$ so that 
\begin{equation}
\tilde{\boldsymbol{\gamma}}^{pr}\tilde{\boldsymbol{\gamma}}_{rq}=\delta_{q}^{p}\,,\quad \boldsymbol{E}_{I}{}^{q}\boldsymbol{E}^{J}{}_{q}=\delta_{I}^{J}\,,\quad \boldsymbol{E}_{I}{}^{p}\boldsymbol{E}^{I}{}_{q}=\delta_{q}^{p}\,,\quad\tilde{\boldsymbol{\gamma}}^{pq}=\delta^{IJ}\boldsymbol{E}_{I}{}^{p}\boldsymbol{E}_{J}{}^{q}\,.
\end{equation}
The quantities $\boldsymbol{M}$, $\boldsymbol{E}^{I}{}_{p}$, $\boldsymbol{E}_{I}{}^{p}$, $\tilde{\boldsymbol{\gamma}}_{pq}$
and $\tilde{\boldsymbol{\gamma}}^{pq}$ are spacetime scalars but they do not depend
explicitly on the spacetime coordinates $x^{\mu}$. Instead, they
depend on the spacetime coordinates $x^{\mu}$ implicitly only through
($\phi^{0}$, $\phi^{r}$).

In order to construct the covariant action of the precursor theory, we define matrix-valued spacetime scalars $\boldsymbol{X}_{I}{}^{J}$ and $\boldsymbol{Y}_{I}{}^{J}$ as 
\begin{equation}
\boldsymbol{X}_{I}{}^{J}\equiv \boldsymbol{e}_{I}{}^{p}\boldsymbol{E}^{J}{}_{p}\,,\quad \boldsymbol{Y}_{I}{}^{J}\equiv \boldsymbol{E}_{I}{}^{p}\boldsymbol{e}^{J}{}_{p}\,.
\end{equation}
As matrices, they are the inverse of each other, since
\begin{equation}
\boldsymbol{X}_{I}{}^{K}\boldsymbol{Y}_{K}{}^{J} = \delta_{I}^{J}\,,\quad \boldsymbol{Y}_{I}{}^{K}\boldsymbol{X}_{K}{}^{J} = \delta_{I}^{J}\,.
\end{equation}

The action of the precursor theory is then written covariantly as
\begin{eqnarray}
S_{\mathrm{pre}} & = & S_{\mathrm{GR}}+\frac{M_{\mathrm{P}}^{2}m^{2}}{2}\,\sum_{n=0}^{4}\int d^{4}x\sqrt{-g}\,c_{n}\mathcal{L}_{n}\,,\\
S_{\mathrm{GR}} & = & \frac{M_{\mathrm{P}}^{2}}{2}\int d^{4}x\sqrt{-g}\mathcal{R}[g_{\mu\nu}]\,,\\
\mathcal{L}_{0} & = & -|\det \boldsymbol{X}|\frac{\boldsymbol{M}}{\boldsymbol{N}}\,,\\
\mathcal{L}_{1} & = & -|\det \boldsymbol{X}|\left(1+\frac{\boldsymbol{M}}{\boldsymbol{N}}\boldsymbol{Y}_{I}{}^{I}\right)\,,\\
\mathcal{L}_{2} & = & -|\det \boldsymbol{X}|\left[\boldsymbol{Y}_{I}{}^{I}+\frac{1}{2}\frac{\boldsymbol{M}}{\boldsymbol{N}}\left(\boldsymbol{Y}_{I}{}^{I}\boldsymbol{Y}_{J}^{\ J}-\boldsymbol{Y}_{I}{}^{J}\boldsymbol{Y}_{J}{}^{I}\right)\right]\,,\\
\mathcal{L}_{3} & = & -\left(\boldsymbol{X}_{I}{}^{I}+\frac{\boldsymbol{M}}{\boldsymbol{N}}\right)\,,\\
\mathcal{L}_{4} & = & -1\,.\label{eqn:action-precursor}
\end{eqnarray}

\section{Quasidilaton extension of precursor theory}\label{sec:quasidilaton-precursor}

\subsection{Covariant action}

Let us introduce a scalar field $\sigma$ and impose the global symmetry,
\begin{equation}
\sigma\to\sigma+\sigma_{0}\,,\quad\phi^{0}\to e^{-(1+\alpha)\sigma_{0}/M_{\mathrm{P}}}\phi^{0}\,,\quad\phi^{p}\to e^{-\sigma_{0}/M_{\mathrm{P}}}\phi^{p}\,,\label{eqn:symmetry}
\end{equation}
where $\sigma_{0}$ is an arbitrary constant, and $\alpha$ is a constant
defining the quasidilaton extension of the minimal theory. We further
impose the symmetry under arbitrary constant shifts of $\phi^{0}$,
arbitrary constant $SO(3)$ rotations in the $3$-dimensional field
space spanned by $\{\phi^{1},\phi^{2},\phi^{3}\}$ and arbitrary constant
translations in the $3$-dimensional field space. Given the assumed symmetry, without further loss of generality, we can then choose 
\begin{equation}
\boldsymbol{M}=1\,,\quad \boldsymbol{E}^{I}{}_{p}=\delta_{p}^{I}\,.\label{eqn:fiducial-shift-SO(3)}
\end{equation}
so that 
\begin{equation}
\boldsymbol{E}_{\ I}^{p}=\delta_{I}^{p}\,,\quad\tilde{\boldsymbol{\gamma}}_{pq}=\delta_{pq}\,,\quad\tilde{\boldsymbol{\gamma}}^{pq}=\delta^{pq}\,,\quad \boldsymbol{X}_{I}{}^{J}=\boldsymbol{e}_{I}{}^{r}\delta^{J}{}_{r}\,,\quad \boldsymbol{Y}_{I}{}^{J}=\delta_{I}^{r}\boldsymbol{e}^{J}{}_{r}\,.
\end{equation}

On adding the scalar $\sigma$ to the MTMG precursor action,
the global symmetry (\ref{eqn:symmetry}) is ensured if we replace
\begin{equation}
\boldsymbol{N}\,,\quad \boldsymbol{X}_{I}{}^{J}\,,\quad \boldsymbol{Y}_{I}{}^{J}\,,
\end{equation}
with 
\begin{equation}
e^{-(1+\alpha)\sigma/M_{\mathrm{P}}}\boldsymbol{N}\,,\quad e^{\sigma/M_{\mathrm{P}}}\boldsymbol{X}_{I}{}^{J}\,,\quad e^{-\sigma/M_{\mathrm{P}}}\boldsymbol{Y}_{I}{}^{J}\,,
\end{equation}
and if we keep 
\begin{equation}
g_{\mu\nu}\,,\quad \boldsymbol{M}=1\,,
\end{equation}
unchanged. Any terms that are invariant under arbitrary constant shifts
of $\sigma$ and that are independent of ($\phi^{0}$, $\phi^{p}$)
can also be added to the action. Therefore, the covariant action of
the quasidilaton extension of the precursor theory is 
\begin{equation}
S_{\mathrm{QDpre}}=S_{\mathrm{GR}}-\frac{\omega}{2}\int dx^{4}\sqrt{-g}g^{\mu\nu}\partial_{\mu}\sigma\partial_{\nu}\sigma+\frac{M_{\mathrm{P}}^{2}m^{2}}{2}\,\sum_{n=0}^{4}\int d^{4}x\sqrt{-g}\,c_{n}\mathcal{L}_{n}^{\mathrm{QD}}\,,
\end{equation}
where 
\begin{eqnarray}
\mathcal{L}_{0}^{\mathrm{QD}} & = & -e^{(4+\alpha)\sigma/M_{\mathrm{P}}}|\det \boldsymbol{X}|\frac{\boldsymbol{M}}{\boldsymbol{N}}\,,\\
\mathcal{L}_{1}^{\mathrm{QD}} & = & -e^{3\sigma/M_{\mathrm{P}}}|\det \boldsymbol{X}|\left(1+\frac{\boldsymbol{M}}{\boldsymbol{N}}e^{\alpha\sigma/M_{\mathrm{P}}}\boldsymbol{Y}_{I}{}^{I}\right)\,,\\
\mathcal{L}_{2}^{\mathrm{QD}} & = & -e^{2\sigma/M_{\mathrm{P}}}|\det \boldsymbol{X}|\left[\boldsymbol{Y}_{I}{}^{I}+\frac{1}{2}\frac{\boldsymbol{M}}{\boldsymbol{N}}e^{\alpha\sigma/M_{\mathrm{P}}}\left(\boldsymbol{Y}_{I}{}^{I}\boldsymbol{Y}_{J}{}^{J}-\boldsymbol{Y}_{I}{}^{J}\boldsymbol{Y}_{J}{}^{I}\right)\right]\,,\\
\mathcal{L}_{3}^{\mathrm{QD}} & = & -e^{\sigma/M_{\mathrm{P}}}\left(\boldsymbol{X}_{I}{}^{I}+\frac{\boldsymbol{M}}{\boldsymbol{N}}e^{\alpha\sigma/M_{\mathrm{P}}}\right)\,,\\
\mathcal{L}_{4}^{\mathrm{QD}} & = & -1\,.\label{eqn:action-precursor-quasidilaton}
\end{eqnarray}
Here, for simplicity we have added only the canonical kinetic term for $\sigma$ (with the dimensionless normalization factor $\omega$) although one could in principle add shift-symmetric Horndeski terms for $\sigma$ to the action without introducing extra propagating degrees of freedom. It is understood that $\boldsymbol{M}$ and $\boldsymbol{E}^{I}{}_{p}$ have been fixed as (\ref{eqn:fiducial-shift-SO(3)}), in particular $\boldsymbol{M}=1$. 

\subsection{Unitary gauge}

On choosing the unitary gauge 
\begin{equation}
\phi^{\mu}=x^{\mu}\,,
\end{equation}
the functions $\boldsymbol{N}$, $\boldsymbol{N}^{p}$, $\boldsymbol{\gamma}_{pq}$ ($p,q,\cdots=1,2,3$) defined in (\ref{eqn:def-N-Np-gammapq}) and (\ref{eqn:def-gammainverse}) are reduced to the lapse $N$, the shift $N^{i}$ and the $3$-dimensional spatial metric $\gamma_{ij}$ ($i,j,\cdots=1,2,3$), which satisfy
\begin{equation}
g_{\mu\nu}dx^{\mu}dx^{\nu}=-N^{2}dt^{2}+\gamma_{ij}(dx^{i}+N^{i}dt)(dx^{j}+N^{j}dt)\,,
\end{equation}
and the set of scalars $\boldsymbol{e}^{I}{}_{p}$ ($I,\cdots=1,2,3$) defined by (\ref{eqn:def-eIp}) and (\ref{eqn:def-einverse}) are reduced to the components of spatial vielbein $e^{I}{}_{i}$ so that 
\begin{equation}
\gamma_{ij}=\delta_{IJ}e^{I}{}_{i}e^{J}{}_{j}\,.
\end{equation}

The action of the precursor theory with the quasidilaton then becomes 
\begin{eqnarray}
S_{\mathrm{QDpre}} & = & \int d^{4}x\,\Bigl\{\frac{M_{\mathrm{P}}^{2}}{2}N\sqrt{\gamma}\,(R[\gamma_{ij}]+K_{ij}K^{ij}-K^{2}\bigr)+\frac{\omega}{2}\sqrt{\gamma}\left[\frac{1}{N}(\dot{\sigma}-N^{i}\partial_{i}\sigma)^{2}-N\gamma^{ij}\partial_{i}\sigma\partial_{j}\sigma\right]\nonumber \\
 &  & -c_{0}\frac{M_{\mathrm{P}}^{2}}{2}m^{2}\sqrt{\tilde{\gamma}}Me^{(4+\alpha)\sigma/M_{\mathrm{P}}}-c_{1}\frac{M_{\mathrm{P}}^{2}}{2}m^{2}\sqrt{\tilde{\gamma}}e^{3\sigma/M_{\mathrm{P}}}(N+Me^{\alpha\sigma/M_{\mathrm{P}}}Y{}_{I}{}^{I})\nonumber \\
 &  & -c_{2}\frac{M_{\mathrm{P}}^{2}}{2}m^{2}\sqrt{\tilde{\gamma}}e^{2\sigma/M_{\mathrm{P}}}\left[NY_{I}{}^{I}+\frac{M}{2}e^{\alpha\sigma/M_{\mathrm{P}}}\,(Y{}_{I}{}^{I}Y{}_{J}{}^{J}-Y{}_{I}{}^{J}Y{}_{J}{}^{I})\right]\nonumber \\
 &  & -c_{3}\frac{M_{\mathrm{P}}^{2}}{2}m^{2}\sqrt{\gamma}e^{\sigma/M_{\mathrm{P}}}(NX{}_{I}{}^{I}+Me^{\alpha\sigma/M_{\mathrm{P}}})-c_{4}\frac{M_{\mathrm{P}}^{2}}{2}m^{2}N\sqrt{\gamma}\Bigr\}\,.\label{eqn:QDpre-unitary}
\end{eqnarray}
Here, we kept again $\sqrt{\tilde{\gamma}}$ and $M$, both of which are
actually $1$, just to make sure that each term has the right density
weight.

\subsection{Hamiltonian analysis of precursor quasidilaton}

\subsubsection{Primary constraints}

Since the graviton mass term is manifestly linear in the lapse and
the shift, we consider $N$ and $N^{i}$ as Lagrange multipliers.
We then have $9$ components of $e^{I}{}_{j}$ and the quasidilaton
scalar $\sigma$ as basic variables. The total number of basic variables
is thus $10$. We define canonical momenta conjugate to them in the
standard way as 
\begin{equation}
\Pi_{I}{}^{j}\equiv\frac{\delta S_{\mathrm{QDpre}}}{\delta\dot{e}^{I}{}_{j}}=2\pi^{jk}\delta_{IJ}e^{J}{}_{k}\,,\label{eqn:piIj-precursor}
\end{equation}
where 
\begin{equation}
\pi^{ij}\equiv\frac{M_{\mathrm{P}}^{2}}{2}\sqrt{\gamma}(K^{ij}-K\gamma^{ij})\,,\quad K_{ij}=\frac{1}{2N}(\dot{\gamma}_{ij}-\mathcal{D}_{i}N_{j}-\mathcal{D}_{j}N_{i})\,,\label{eqn:piK-precursor}
\end{equation}
and 
\begin{equation}
\pi_{\sigma}\equiv\frac{\delta S_{\mathrm{QDpre}}}{\delta\dot{\sigma}}=\frac{\omega\sqrt{\gamma}}{N}(\dot{\sigma}-N^{i}\partial_{i}\sigma)\,.\label{eqn:pisigma-precursor}
\end{equation}

The fact that $K^{ij}$ is symmetric leads to the following $3$ primary
constraints 
\begin{equation}
{\cal P}_{[IJ]}\approx0\,,\label{eqn:primaryconstraint}
\end{equation}
where 
\begin{equation}
{\cal P}_{[IJ]}\equiv\Pi_{[I}{}^{k}\delta_{J]K}e^{K}{}_{k}\,,
\end{equation}
and indices between the square brackets are antisymmetrized as $A_{[ab]}=A_{ab}-A_{ba}$.
The remaining $10-3=7$ relations between the canonical momenta and
the time derivative of the basic variables can be inverted as 
\begin{equation}
\delta_{IJ}\dot{e}_{(i}^{I}e_{j)}^{J}=NK_{ij}+\frac{1}{2}(\mathcal{D}_{i}N_{j}+\mathcal{D}_{j}N_{i})\,,\quad K_{ij}=\frac{1}{M_{\mathrm{P}}^{2}\sqrt{\gamma}}\left[\gamma_{k(i}\gamma_{j)l}\Pi_{I}^{\ k}\delta^{IJ}e_{J}^{\ l}-\frac{1}{2}\gamma_{kl}\Pi_{K}{}^{k}\delta^{KL}e_{L}{}^{l}\gamma_{ij}\right]\,,
\end{equation}
and 
\begin{equation}
\dot{\sigma}=\frac{N}{\omega\sqrt{\gamma}}\pi_{\sigma}+N^{i}\partial_{i}\sigma\,.
\end{equation}
Thus there are no more primary constraints associated with (\ref{eqn:piIj-precursor})
and (\ref{eqn:pisigma-precursor}).

The Hamiltonian of the quasidilaton precursor theory, together with
the primary constraints, is 
\begin{equation}
\bar{H}_{\mathrm{QDpre}}^{(1)}=\int d^{3}x\,[-N\mathcal{R}_{0}-N^{i}\mathcal{R}_{i}+\frac{M_{\mathrm{P}}^{2}}{2}m^{2}M\mathcal{H}_{1}+\alpha_{MN}\mathcal{P}^{[MN]}]\,,
\end{equation}
where 
\begin{eqnarray*}
\mathcal{R}_{0} & = & \mathcal{R}_{0}^{\mathrm{GR}}-\frac{M_{\mathrm{P}}^{2}}{2}m^{2}\mathcal{H}_{0}\,,\\
\mathcal{R}_{0}^{\mathrm{GR}} & = & \frac{M_{\mathrm{P}}^{2}}{2}\sqrt{\gamma}\,R[\gamma]-\frac{2}{M_{\mathrm{P}}^{2}}\frac{1}{\sqrt{\gamma}}\left(\gamma_{il}\gamma_{jk}-\frac{1}{2}\gamma_{ij}\gamma_{kl}\right)\pi^{ij}\pi^{kl}-\mathcal{H}_{\sigma}\,,\\
\mathcal{H}_{\sigma} & = & \frac{1}{2\omega}\frac{1}{\sqrt{\gamma}}\pi_{\sigma}^{2}+\frac{\omega}{2}\sqrt{\gamma}\gamma^{ij}\partial_{i}\sigma\partial_{j}\sigma\,,\\
\mathcal{R}_{i} & = & \mathcal{R}_{i}^{\mathrm{GR}}=2\sqrt{\gamma}\gamma_{ik}\mathcal{D}_{j}\left(\frac{\pi^{kj}}{\sqrt{\gamma}}\right)-\pi_{\sigma}\partial_{i}\sigma\,,\\
\mathcal{H}_{0} & = & \left[\sqrt{\tilde{\gamma}}\left(c_{1}e^{3\sigma/M_{\mathrm{P}}}+c_{2}e^{2\sigma/M_{\mathrm{P}}}\,Y{}_{I}{}^{I}\right)+\sqrt{\gamma}(c_{3}e^{\sigma/M_{\mathrm{P}}}\,X{}_{I}{}^{I}+c_{4})\right]\,,\\
\mathcal{H}_{1} & = & e^{\alpha\sigma/M_{\mathrm{P}}}\left\{ \sqrt{\tilde{\gamma}}\left[c_{0}e^{4\sigma/M_{\mathrm{P}}}+c_{1}e^{3\sigma/M_{\mathrm{P}}}Y{}_{I}{}^{I}+\frac{c_{2}}{2}e^{2\sigma/M_{\mathrm{P}}}\,(Y{}_{I}{}^{I}Y{}_{J}{}^{J}-Y{}_{I}{}^{J}Y{}_{J}{}^{I})\right]+c_{3}\sqrt{\gamma}e^{\sigma/M_{\mathrm{P}}}\right\} \,,\\
\mathcal{P}^{[MN]} & = & e^{M}{}_{j}\,\Pi^{j}{}_{I}\delta^{IN}-e^{N}{}_{j}\,\Pi^{j}{}_{I}\,\delta^{IM}\,,
\end{eqnarray*}
$\mathcal{D}_{j}$ is the spatial covariant derivative compatible
with $\gamma_{ij}$, $\sqrt{\gamma}=\sqrt{\det\gamma_{ij}}$, $\sqrt{\tilde{\gamma}}=\sqrt{\det\tilde{\gamma}_{ij}}=1$,
$M=1$ and $\alpha_{MN}$ (antisymmetric) are $3$ Lagrange multipliers.

The Hamiltonian is manifestly linear in the lapse $N$ and the shift
$N^{i}$ and does not contain their time derivatives. Thus, as already
stated, we consider $N$ and $N^{i}$ as Lagrange multipliers. Correspondingly,
we have the following primary constraints in addition to (\ref{eqn:primaryconstraint}):
\begin{equation}
\mathcal{R}_{0}\approx0\,,\quad\mathcal{R}_{i}\approx0\,.
\end{equation}

\subsubsection{Secondary constraints and total Hamiltonian}

In order to implement the conservation in time of the primary constraints,
we need the following Poisson brackets to vanish 
\begin{eqnarray}
\dot{\mathcal{P}}^{[MN]} & = & \{{\mathcal{P}^{[MN]},\bar{H}_{{\rm QDpre}}^{(1)}}\}\approx0\,,\label{eq:dotP}\\
\dot{\mathcal{R}}_{0} & = & \{{\mathcal{R}_{0},\bar{H}_{{\rm QDpre}}^{(1)}}\}\approx0\,,\label{eq:dotR0}\\
\dot{\mathcal{R}}_{i} & = & \{{\mathcal{R}_{i},\bar{H}_{{\rm QDpre}}^{(1)}}\}\approx0\,.\label{eq:dotRi}
\end{eqnarray}
Even though we have chosen the unitary gauge, we have omitted the partial time derivative of $\mathcal{R}_0$ in Eq.\ (\ref{eq:dotR0}) because, for our choice of the fiducial vielbein/metric (\ref{eqn:fiducial-shift-SO(3)}), we have $\partial\mathcal{R}_{0}/\partial t=0$. Then Eq.\ (\ref{eq:dotP})
leads to three new secondary constraints, namely 
\begin{equation}
Y^{[MN]}\approx0\,,
\end{equation}
where we have defined 
\begin{equation}
Y^{MN}=\delta^{ML}Y_{L}{}^{N}\,.
\end{equation}
This set of secondary constraints fixes $Y^{MN}$ to be symmetric.

Since 
\begin{eqnarray}
\{\mathcal{R}_{0}(x),\mathcal{R}_{0}(y)\} & \approx & 0\,,\\
\{\mathcal{R}_{i}(x),\mathcal{R}_{j}(y)\} & \approx & 0\,,\\
\{\mathcal{R}_{0}(x),\mathcal{R}_{i}(y)\} & \not\approx & 0\,,
\end{eqnarray}
we can use Eq.\ (\ref{eq:dotR0}) to find the expression of one of
the components of $N^{i}$ (say $N^{i=3}$) in terms of the other
variables. For the same reason we can solve one of the three equations (\ref{eq:dotRi})
(say for $i=3$) for the lapse variable $N$. Therefore the remaining
two equations (\ref{eq:dotRi}) give rise to two secondary constraints,
(say $\dot{\mathcal{R}}_{1}\approx0$ and $\dot{\mathcal{R}}_{2}\approx0$
after solving $\dot{\mathcal{R}}_{3}\approx0$ with respect to one
of Lagrange multipliers). On naming these two constraints as $\tilde{\mathcal{C}}_{\tau}$
($\tau=1,2$), then we have the total Hamiltonian 
\begin{equation}
\bar{H}_{\mathrm{QDpre}}^{(2)}=\int d^{3}x\,\left[-N\mathcal{R}_{0}-N^{i}\mathcal{R}_{i}+\frac{M_{\mathrm{P}}^{2}}{2}m^{2}M\mathcal{H}_{1}+\alpha_{MN}\mathcal{P}^{[MN]}+\beta_{MN}Y^{[MN]}+\tilde{\lambda}^{\tau}\tilde{\mathcal{C}}_{\tau}\right]\,.\label{eq:Htot2nd}
\end{equation}
Any further time derivatives of the constraints do not lead to any
new (tertiary) constraints, therefore Eq.\ (\ref{eq:Htot2nd}) represents
the total Hamiltonian.

\subsubsection{Number of physical degrees of freedom in precursor theory}

It is straightforward to show that the determinant of the $12\times12$
matrix made of the Poisson brackets among $12$ constraints is nonvanishing.
This implies that the $12$ constraints are independent second-class
constraints and that the consistency of them with the time evolution
uniquely determines all Lagrange multipliers without generating additional
constraints. Since each of these $12$ second-class constraints removes
one single degree of freedom in the phase space, we finally have $\frac{1}{2}(10\times2-12)=4$
physical degrees of freedom on a generic background at nonlinear level.

\section{Hamiltonian of minimal quasidilaton theory}

\label{sec:hamiltonian}

We have seen that, besides $Y^{[MN]}\approx0$, the precursor quasidilaton
theory possesses the two secondary constraints $\tilde{\mathcal{C}}_{\tau}$
($\tau=1,2$), which are two linear combinations of the three quantities
$\mathcal{C}_{i}$ ($i=1,2,3$) defined as follows 
\[
\{\mathcal{R}_{i}^{\mathrm{GR}},H_{1}\}\approx\frac{M_{\mathrm{P}}^{2}}{2}\mathcal{C}_{i}\,,
\]
where 
\begin{equation}
H_{1}=\frac{M_{\mathrm{P}}^{2}}{2}m^{2}\int d^{3}xM\mathcal{H}_{1}\,.
\end{equation}
The minimal quasidilaton theory is defined by imposing the four constraints
\begin{equation}
\mathcal{C}_{0}\approx0,\quad\mathcal{C}_{i}\approx0\,,\label{eq:constrTot}
\end{equation}
where 
\[
\{\mathcal{R}_{0}^{\mathrm{GR}},H_{1}\}\approx\frac{M_{\mathrm{P}}^{2}}{2}\mathcal{C}_{0}\,.
\]
Since $\tilde{\mathcal{C}}_{\tau}$ ($\tau=1,2$) are linear combinations
of $\mathcal{C}_{i}$, only two constraints among the four in (\ref{eq:constrTot})
are independent new constraints. Therefore, the minimal quasidilaton
theory is defined by the Hamiltonian 
\begin{equation}
H=\int d^{3}x\left[-N\mathcal{R}_{0}^{\mathrm{GR}}-N^{i}\mathcal{R}_{i}^{\mathrm{GR}}+\frac{M_{\mathrm{P}}^{2}}{2}m^{2}(N\mathcal{H}_{0}+M\mathcal{H}_{1})+\frac{M_{\mathrm{P}}^{2}}{2}(\lambda\mathcal{C}_{0}+\lambda^{i}\mathcal{C}_{i})+\alpha_{MN}\mathcal{P}^{MN}+\beta_{MN}Y^{[MN]}\right]\,,\label{eq:HtotMin}
\end{equation}
where 
\begin{eqnarray*}
\mathcal{R}_{0}^{\mathrm{GR}} & = & \frac{M_{\mathrm{P}}^{2}}{2}\sqrt{\gamma}\,R[\gamma]-\frac{2}{M_{\mathrm{P}}^{2}}\frac{1}{\sqrt{\gamma}}\left(\gamma_{nl}\gamma_{mk}-\frac{1}{2}\gamma_{nm}\gamma_{kl}\right)\pi^{nm}\pi^{kl}-\mathcal{H}_{\sigma}\,,\\
\mathcal{H}_{\sigma} & = & \frac{1}{2\omega}\frac{1}{\sqrt{\gamma}}\pi_{\sigma}^{2}+\frac{\omega}{2}\sqrt{\gamma}\gamma^{ij}\partial_{i}\sigma\partial_{j}\sigma\,,\\
\mathcal{R}_{i}^{\mathrm{GR}} & = & 2\sqrt{\gamma}\gamma_{ik}\mathcal{D}_{j}\left(\frac{\pi^{kj}}{\sqrt{\gamma}}\right)-\pi_{\sigma}\partial_{i}\sigma\,,\\
\mathcal{H}_{0} & = & \left[\sqrt{\tilde{\gamma}}\left(c_{1}e^{3\sigma/M_{\mathrm{P}}}+c_{2}e^{2\sigma/M_{\mathrm{P}}}\,Y{}_{I}{}^{I}\right)+\sqrt{\gamma}(c_{3}e^{\sigma/M_{\mathrm{P}}}\,X{}_{I}{}^{I}+c_{4})\right]\,,\\
\mathcal{H}_{1} & = & e^{\alpha\sigma/M_{\mathrm{P}}}\left\{ \sqrt{\tilde{\gamma}}\left[c_{0}e^{4\sigma/M_{\mathrm{P}}}+c_{1}e^{3\sigma/M_{\mathrm{P}}}Y{}_{I}{}^{I}+\frac{c_{2}}{2}e^{2\sigma/M_{\mathrm{P}}}\,(Y{}_{I}{}^{I}Y{}_{J}{}^{J}-Y{}_{I}{}^{J}Y{}_{J}{}^{I})\right]+c_{3}\sqrt{\gamma}e^{\sigma/M_{\mathrm{P}}}\right\} \,,\\
\mathcal{P}^{[MN]} & = & e^{M}{}_{j}\,\Pi^{j}{}_{I}\delta^{IN}-e^{N}{}_{j}\,\Pi^{j}{}_{I}\,\delta^{IM}\,,\\
Y^{[MN]} & = & \delta^{MI}Y_{I}{}^{N}-\delta^{NI}Y_{I}{}^{M}\,,
\end{eqnarray*}
and 
\begin{eqnarray}
\mathcal{C}_{0} & = & m^{2}M\left[\frac{1}{2}W_{I}{}^{J}(\gamma_{ik}E_{J}{}^{k}e^{I}{}_{j}+\gamma_{jk}E_{J}{}^{k}e^{I}{}_{i}-\gamma_{ij}Y_{J}{}^{I})\frac{2 \pi^{ij}}{M^2_\mathrm{P}}+\frac{1}{\omega}\frac{1}{\sqrt{\gamma}}\frac{\partial\mathcal{H}_{1}}{\partial\sigma}\pi_{\sigma}\right]\,,\nonumber \\
\mathcal{C}_{i} & = & m^{2}\left[-\sqrt{\gamma}\mathcal{D}^{j}\!\left(MW_{I}{}^{J}Y_{J}{}^{K}\delta_{KL}e^{I}{}_{i}e^{L}{}_{j}\right)+M\frac{\partial\mathcal{H}_{1}}{\partial\sigma}\partial_{i}\sigma\right]\,.
\end{eqnarray}
Here we have defined 
\begin{eqnarray}
W_{I}{}^{J} & = & e^{\alpha\sigma/M_{\mathrm{P}}}\left\{ \frac{\sqrt{\tilde{\gamma}}}{\sqrt{\gamma}}\left[c_{1}e^{3\sigma/M_{\mathrm{P}}}\delta_{I}^{J}+c_{2}e^{2\sigma/M_{\mathrm{P}}}(Y_{K}{}^{K}\delta_{I}^{J}-Y_{I}{}^{J})\right]+c_{3}e^{\sigma/M_{\mathrm{P}}}X_{I}{}^{J}\right\} \,,\nonumber \\
\frac{\partial\mathcal{H}_{1}}{\partial\sigma} & = & \frac{e^{\alpha\sigma/M_{\mathrm{P}}}}{M_{\mathrm{P}}}\left\{ \sqrt{\tilde{\gamma}}\left[(4+\alpha)c_{0}e^{4\sigma/M_{\mathrm{P}}}+(3+\alpha)c_{1}e^{3\sigma/M_{\mathrm{P}}}Y{}_{I}{}^{I}\right.\right.\nonumber \\
 &  & \left.\left.+\frac{2+\alpha}{2}c_{2}e^{2\sigma/M_{\mathrm{P}}}\,(Y{}_{I}{}^{I}Y{}_{J}{}^{J}-Y{}_{I}{}^{J}Y{}_{J}{}^{I})\right]+(1+\alpha)c_{3}\sqrt{\gamma}e^{\sigma/M_{\mathrm{P}}}\right\} \,.
\end{eqnarray}
Again, in the above expression we kept $\sqrt{\tilde{\gamma}}$ and $M$, both of which are actually $1$ (see (\ref{eqn:fiducial-shift-SO(3)})), just to make sure that each term has the right density weight. The fact that $M$ ($=1$) is constant implies that $\mathcal{C}_{i}$ is
independent of $c_{3}$.

The main difference between the two Hamiltonians in equations (\ref{eq:HtotMin})
and (\ref{eq:Htot2nd}) consists in the presence of the four constraints
($\mathcal{C}_{0}$, $\mathcal{C}_{i}$) rather than the two constraints
$\tilde{\mathcal{C}}_{\tau}$. Furthermore the constraints ($\mathcal{C}_{0}$,
$\mathcal{C}_{i}$) are the time derivative of the primary constraints
with respect to $H_{1}$ (and not $H$, although $H\approx H_{1}$).

\subsection{Number of physical degrees of freedom in minimal quasidilaton theory}

\label{subsec:minimaltheory-nopdf}

Having added the extra two constraints, we now have $14$ constraints
in the $10\times2=20$ dimensional phase space. Thus the number of
dimensions of the physical phase space is less than or equal to $20-14=6$,
where the equality holds if all $14$ constraints are second-class
and if there is no more constraint. Therefore, we conclude that $(\mbox{number of d.o.f.})\leq\frac{1}{2}\cdot6=3$
at the fully nonlinear level. On the other hand, in Sec.\ \ref{sec:stability}
we shall explicitly show that cosmological perturbations around de
Sitter backgrounds contain two tensor modes (gravitational waves)
and one scalar mode (quasidilaton perturbation) at the linear level,
meaning that $(\mbox{number of d.o.f.})\geq3$ at the nonlinear level.
Combining the two inequalities we conclude that $(\mbox{number of d.o.f.})=3$.

One can reach the same conclusion also in a more formal way. Since
the actual calculation is somehow cumbersome, we shall simply give
a brief outline. What we need to show is that the consistency of the
$14$ constraints with the time evolution does not lead to additional
constraints but simply determines all Lagrange multipliers. For this
purpose it is necessary and sufficient to show that the determinant
of the matrix $\{\mathcal{Z}^{\sigma_{1}}(x),\mathcal{Z}^{\sigma_{2}}(y)\}$
is non-vanishing, where $\mathcal{Z}^{\sigma}(x)$ ($\sigma=1,\cdots,14$)
represents the $14$ constraints. In other words, we need to show
that, for a vector field $v_{\sigma}$, the set of $14$ equations
\begin{equation}
\int dy\{\mathcal{Z}^{\sigma_{1}}(x),\mathcal{Z}^{\sigma_{2}}(y)\}v_{\sigma_{2}}(y)\approx0\,,
\end{equation}
has the unique solution $v_{\sigma}=0$. Once this proposition is
proved, we can conclude that all the $14$ constraints are independent
second-class constraints and that the consistency of them with the
time evolution does not lead to additional constraints. Since we have
$14$ second-class constraints in the $10\times2=20$ dimensional
phase space, the number of physical degrees of freedom in this theory
is $\frac{1}{2}\cdot(10\times2-14)=3$ at fully nonlinear level.

\section{Lagrangian of minimal quasidilaton theory}

\label{sec:lagrangian}

\subsection{Vielbein formulation of minimal quasidilaton theory}

The Hamiltonian equation of motion for $e^{I}{}_{j}$ can be inverted
to express $\pi^{ij}$ and $\Pi_{I}{}^{j}$ in terms of the extrinsic
curvature as 
\begin{equation}
\frac{2}{M_{\mathrm{P}}^{2}}\frac{\pi^{ij}}{\sqrt{\gamma}}=K^{ij}-K\gamma^{ij}-\frac{m^{2}}{4}\frac{M}{N}\lambda\Theta^{ij}\,,\label{eqn:piK}
\end{equation}
and 
\begin{equation}
\Pi_{I}{}^{j}=2\pi^{jk}\delta_{IJ}e^{J}{}_{k}\,,\label{eqn:piIj}
\end{equation}
where 
\begin{equation}
\Theta^{ij}=W_{I}{}^{J}\delta^{IK}(e_{K}{}^{i}E_{J}{}^{j}+e_{K}{}^{j}E_{J}{}^{i})\,.
\end{equation}
Equivalently, 
\begin{equation}
\Theta^{i}{}_{j}=W_{I}{}^{J}(\delta^{IK}e_{K}{}^{i}Y_{J}{}^{L}\delta_{LM}e^{M}{}_{j}+e^{I}{}_{j}E_{J}{}^{i}).
\end{equation}
The relation between $\pi_{\sigma}$ and $\dot{\sigma}$ derived from
the Hamiltonian equation of motion is 
\begin{equation}
\pi_{\sigma}=\frac{\omega\sqrt{\gamma}}{N}(\dot{\sigma}-N^{i}\partial_{i}\sigma)-\frac{M_{\mathrm{P}}^{2}}{2}m^{2}\lambda\frac{M}{N}\frac{\partial\mathcal{H}_{1}}{\partial\sigma}\,.\label{eqn:pisigma-minimal}
\end{equation}
What is important here is that the relations (\ref{eqn:piK}) and
(\ref{eqn:pisigma-minimal}) in the minimal quasidilaton theory differ
from the corresponding relations (\ref{eqn:piK-precursor}) and (\ref{eqn:pisigma-precursor})
in the precursor quasidilaton theory. This difference stems from the dependence on the canonical momenta included in the additional constraints.

Hence the action of the theory is 
\begin{equation}
S=\int dt\left[\int d^{3}x\left(\Pi_{I}{}^{j}\dot{e}^{I}{}_{j}+\pi_{\sigma}\dot{\sigma}\right)-\left(H\ \mathrm{with}\ \alpha_{MN}=\beta_{MN}=0\right)\right],
\end{equation}
where we have dropped $\alpha_{MN}\mathcal{P}^{MN}$ and $\beta_{MN}Y^{[MN]}$
from the Hamiltonian as they will automatically come out (since $\Theta^{ij}$
is defined as a symmetric tensor, and as we shall explicitly see below)
and it is understood that $\pi^{ij}$ and $\Pi_{I}{}^{j}$ are expressed
in terms of the extrinsic curvature using the above formulas. Explicitly,
\begin{eqnarray}
S & = & S_{\mathrm{QDpre}}-\frac{M_{\mathrm{P}}^{2}}{2}\int d^{4}xN\sqrt{\gamma}\left(\frac{m^{2}}{4}\frac{M}{N}\lambda\right)^{2}\left[\left(\gamma_{ik}\gamma_{jl}-\frac{1}{2}\gamma_{ij}\gamma_{kl}\right)\Theta^{ij}\Theta^{kl}+\frac{M_{\mathrm{P}}^{2}}{2}\frac{8}{\omega}\left(\frac{1}{\sqrt{\gamma}}\frac{\partial\mathcal{H}_{1}}{\partial\sigma}\right)^{2}\right]\nonumber \\
 &  & -\frac{M_{\mathrm{P}}^{2}}{2}\int d^{4}x\left(\lambda\mathcal{C}_{0}+\lambda^{i}\mathcal{C}_{i}\right)\nonumber \\
 & = & S_{\mathrm{QDpre}}+\frac{M_{\mathrm{P}}^{2}}{2}\int d^{4}xN\sqrt{\gamma}\left(\frac{m^{2}}{4}\frac{M}{N}\lambda\right)^{2}\left[\left(\gamma_{ik}\gamma_{jl}-\frac{1}{2}\gamma_{ij}\gamma_{kl}\right)\Theta^{ij}\Theta^{kl}+\frac{M_{\mathrm{P}}^{2}}{2}\frac{8}{\omega}\left(\frac{1}{\sqrt{\gamma}}\frac{\partial\mathcal{H}_{1}}{\partial\sigma}\right)^{2}\right]\nonumber \\
 &  & -\frac{M_{\mathrm{P}}^{2}}{2}\int d^{4}x\left(\lambda\bar{\mathcal{C}}_{0}+\lambda^{i}\mathcal{C}_{i}\right),
\end{eqnarray}
where $S_{\mathrm{QDpre}}$ is the unitary-gauge action for the precursor quasidilaton theory given in (\ref{eqn:QDpre-unitary}). It is understood that $\mathcal{C}_{0}$ is now defined as
\begin{equation}
\mathcal{C}_{0}=m^{2}M\left[\sqrt{\gamma}W_{I}{}^{J}\left(\gamma_{ik}E_{J}{}^{k}e^{I}{}_{j}-\frac{1}{2}\gamma_{ij}Y_{J}{}^{I}\right)\left(K^{ij}-K\gamma^{ij}-\frac{m^{2}}{4}\frac{M}{N}\lambda\Theta^{ij}\right)+\frac{\partial\mathcal{H}_{1}}{\partial\sigma}\frac{1}{N}\left(\dot{\sigma}-N^{i}\partial_{i}\sigma-\frac{m^{2}\lambda}{\omega\sqrt{\gamma}}\frac{\partial\mathcal{H}_{1}}{\partial\sigma}\right)\right]\,,
\end{equation}
while $\mathcal{C}_{i}$, $\mathcal{P}^{MN}$ and $Y^{[MN]}$ are
defined as before. Finally, $\bar{\mathcal{C}}_{0}$ is defined as
\begin{equation}
\bar{\mathcal{C}}_{0}\equiv\mathcal{C}_{0}|_{\lambda=0}=m^{2}M\left[\sqrt{\gamma}W_{I}{}^{J}\left(\gamma_{ik}E_{J}{}^{k}e^{I}{}_{j}-\frac{1}{2}\gamma_{ij}Y_{J}{}^{I}\right)\left(K^{ij}-K\gamma^{ij}\right)+\frac{\partial\mathcal{H}_{1}}{\partial\sigma}\frac{1}{N}\left(\dot{\sigma}-N^{i}\partial_{i}\sigma\right)\right]\,.
\end{equation}

As a consistency check, let us calculate the Hamiltonian of the system
defined by the action and compare it with the Hamiltonian defined
in the previous section. The system has the following primary constraints
\begin{equation}
\pi_{N}=0\,,\quad\pi_{i}=0\,,\quad\pi^{\lambda}=0\,,\quad\pi_{i}^{\lambda}=0\,,\quad\mathcal{P}^{[MN]}=0\,,
\end{equation}
where $\pi_{N}$, $\pi_{i}$, $\pi^{\lambda}$ and $\pi_{i}^{\lambda}$
are canonical momenta conjugate to $N$, $N^{i}$, $\lambda$ and
$\lambda^{i}$, respectively, and $\mathcal{P}^{[MN]}$ is defined
in the previous section. The canonical momenta conjugate to $e^{I}{}_{j}$
is then given precisely by (\ref{eqn:piIj}). The Hamiltonian is then
\begin{equation}
\tilde{H}=H+\int d^{3}x\left(\Lambda^{N}\pi_{N}+\Lambda^{i}\pi_{i}+\Lambda_{\lambda}\pi^{\lambda}+\Lambda_{\lambda}^{i}\pi_{i}^{\lambda}\right),\label{eqn:tildeH}
\end{equation}
where $H$ (with $\alpha_{MN}P^{[MN]}$ and $\beta_{MN}Y^{[MN]}$
included) was defined in the previous section and $Y^{[MN]}$ has
been added to the Hamiltonian as a solution to the secondary constraint
associated with the primary constraint $P^{[MN]}=0$. Since $H$ depends
linearly on $N$, $N^{i}$, $\lambda$ and $\lambda^{i}$, it is obvious
that $\pi_{N}=0$, $\pi_{i}=0$, $\pi^{\lambda}=0$ and $\pi_{i}^{\lambda}=0$
are first-class. We can then safely downgrade $N$, $N^{i}$, $\lambda$
and $\lambda^{i}$ to Lagrange multipliers, and drop $\pi_{N}$, $\pi_{i}$,
$\pi^{\lambda}$ and $\pi_{i}^{\lambda}$ from the phase space variables.
After that, the Hamiltonian $\tilde{H}$ in (\ref{eqn:tildeH}) becomes
manifestly equivalent to $H$ defined in the previous section.

\subsection{Metric formulation of minimal quasidilaton theory}

Consider the spatial tensor $\mathcal{K}^{i}{}_{j}$ defined so that 
\begin{equation}
\mathcal{K}^{i}{}_{l}\mathcal{K}^{l}{}_{j}=\tilde{\gamma}^{il}\gamma_{lj}\,,
\end{equation}
and define its inverse, $\mathfrak{K}^{i}{}_{j}$, as 
\begin{equation}
\mathfrak{K}^{i}{}_{l}\mathcal{K}^{l}{}_{j}=\delta^{i}{}_{j}\,.
\end{equation}
In terms of the vielbein we can write 
\begin{eqnarray}
\mathcal{K}^{i}{}_{j} & = & E_{L}{}^{i}e^{L}{}_{j}\,,\\
\mathfrak{K}^{i}{}_{j} & = & e_{L}{}^{i}E^{L}{}_{j}\,.
\end{eqnarray}
In the metric formalism, provided that $Y_{I}{}^{J}=E_{I}{}^{i}e^{J}{}_{i}$
is symmetric, we have 
\begin{eqnarray}
\mathcal{K}^{i}{}_{j} & \equiv & \left(\sqrt{\tilde{\gamma}^{-1}\gamma}\right)^{i}{}_{j}\,,\\
\mathfrak{K}^{i}{}_{j} & \equiv & \left(\sqrt{\gamma^{-1}\tilde{\gamma}}\right)^{i}{}_{j}\,,\\
\mathfrak{K}^{i}{}_{l}\mathcal{K}^{l}{}_{j} & = & \delta^{i}{}_{j}=\mathcal{K}^{i}{}_{l}\mathfrak{K}^{l}{}_{j}\,.
\end{eqnarray}
Let us build the following tensor 
\begin{equation}
\Theta^{ij}=e^{\alpha\sigma/M_{\mathrm{P}}}\left[\frac{\sqrt{\tilde{\gamma}}}{\sqrt{\gamma}}\{c_{1}e^{3\sigma/M_{\mathrm{P}}}(\gamma^{il}\mathcal{K}^{j}{}_{l}+\gamma^{jl}\mathcal{K}^{i}{}_{l})+c_{2}e^{2\sigma/M_{\mathrm{P}}}[\mathcal{K}(\gamma^{il}\mathcal{K}^{j}{}_{l}+\gamma^{jl}\mathcal{K}^{i}{}_{l})-2\tilde{\gamma}^{ij}]\}+2c_{3}e^{\sigma/M_{\mathrm{P}}}\gamma^{ij}\right]\,,
\end{equation}
and the following spatial scalar density 
\begin{eqnarray}
\frac{\partial\mathcal{H}_{1}}{\partial\sigma} & = & \frac{e^{\alpha\sigma/M_{\mathrm{P}}}}{M_{\mathrm{P}}}\left\{ \sqrt{\tilde{\gamma}}\left[(4+\alpha)c_{0}e^{4\sigma/M_{\mathrm{P}}}+(3+\alpha)c_{1}e^{3\sigma/M_{\mathrm{P}}}\mathcal{K}^{i}{}_{i}\right.\right.\nonumber \\
 &  & \left.\left.+\frac{2+\alpha}{2}c_{2}e^{2\sigma/M_{\mathrm{P}}}\,(\mathcal{K}^{i}{}_{i}\mathcal{K}^{j}{}_{j}-\mathcal{K}^{i}{}_{j}\mathcal{K}^{j}{}_{i})\right]+(1+\alpha)c_{3}\sqrt{\gamma}e^{\sigma/M_{\mathrm{P}}}\right\} \,.
\end{eqnarray}
We further define the four constraints imposed on the system in order
to reduce the degrees of freedom: 
\begin{eqnarray}
\bar{\mathcal{C}}_{0} & = & m^{2}\left\{ \frac{1}{2}M\sqrt{\gamma}\left(\gamma_{ik}\gamma_{jl}-\frac{1}{2}\gamma_{ij}\gamma_{kl}\right)\Theta^{kl}\left(K^{ij}-K\gamma^{ij}\right)+\frac{\partial\mathcal{H}_{1}}{\partial\sigma}\frac{M}{N}\left(\dot{\sigma}-N^{i}\partial_{i}\sigma\right)\right\} \,,\\
C_{i} & = & m^{2}\left[-\sqrt{\gamma}\mathcal{D}_{j}\left(Me^{\alpha\sigma/M_{{\rm P}}}\left\{ \frac{\sqrt{\tilde{\gamma}}}{\sqrt{\gamma}}\left[\frac{1}{2}\,(c_{1}e^{3\sigma/M_{{\rm P}}}+c_{2}e^{2\sigma/M_{{\rm P}}}\mathcal{K})\left(\mathcal{K}^{j}{}_{i}+\gamma^{jk}\mathcal{K}^{l}{}_{k}\gamma_{li}\right)-c_{2}e^{2\sigma/M_{{\rm P}}}\tilde{\gamma}^{jk}\gamma_{ki}\right]\right.\right.\right.\nonumber \\
 &  & {}+\left.\left.\left.c_{3}e^{\sigma/M_{{\rm P}}}\delta^{j}{}_{i}\right\} \right)+M\,\frac{\partial\mathcal{H}_{1}}{\partial\sigma}\,\partial_{i}\sigma\right],
\end{eqnarray}
where $K^{ij}$ is the extrinsic curvature and $\mathcal{K}$ represents
$\mathcal{K}^{i}{}_{i}$. As already stated in Sec.\ \ref{sec:hamiltonian},
the fact that $M$ ($=1$) is constant implies that $\mathcal{C}_{i}$ is
independent of $c_{3}$. The following is the action of the minimal quasidilaton theory written in the metric formalism:
\begin{eqnarray}
S & = & S_{\mathrm{QDpre}}+\frac{M_{\mathrm{P}}^{2}}{2}\int d^{4}xN\sqrt{\gamma}\left(\frac{m^{2}}{4}\,\frac{M}{N}\,\lambda\right)^{\!2}\left[\left(\gamma_{ik}\gamma_{jl}-\frac{1}{2}\gamma_{ij}\gamma_{kl}\right)\Theta^{ij}\Theta^{kl}+\frac{M_{\mathrm{P}}^{2}}{2}\frac{8}{\omega}\left(\frac{1}{\sqrt{\gamma}}\frac{\partial\mathcal{H}_{1}}{\partial\sigma}\right)^{2}\right]\nonumber \\
 &  & {}-\frac{M_{\mathrm{P}}^{2}}{2}\int d^{4}x\left(\lambda\bar{\mathcal{C}}_{0}+\lambda^{i}\,\mathcal{C}_{i}\right)\,.
\end{eqnarray}
As it is well known, in the 3+1 formalism, it is possible to write
the action of general relativity as 
\begin{equation}
S_{\mathrm{GR}}=\frac{M_{\mathrm{P}}^{2}}{2}\,\int d^{4}xN\sqrt{\gamma}\,[{}^{(3)}R+K^{ij}K_{ij}-K^{2}]\,,
\end{equation}
where 
\begin{eqnarray}
K_{ij} & = & \frac{1}{2N}\,(\dot{\gamma}_{ij}-\mathcal{D}_{i}N_{j}-\mathcal{D}_{j}N_{i})\,,\\
K & = & \gamma^{ij}K_{ij}\,.
\end{eqnarray}
Therefore, we have 
\begin{eqnarray}
S_{\mathrm{QDpre}} & = & S_{\mathrm{GR}}+\frac{\omega}{2}\int d^{4}x\sqrt{\gamma}\left[\frac{1}{N}(\dot{\sigma}-N^{i}\partial_{i}\sigma)^{2}-N\gamma^{ij}\partial_{i}\sigma\partial_{j}\sigma\right]+\frac{M_{\mathrm{P}}^{2}}{2}\,\sum_{i=0}^{4}\int d^{4}x\mathcal{S}_{i}\,,\\
\mathcal{S}_{0} & = & -m^{2}c_{0}e^{(4+\alpha)\sigma/M_{\mathrm{P}}}\,\sqrt{\tilde{\gamma}}\,M\,,\\
\mathcal{S}_{1} & = & -m^{2}c_{1}e^{3\sigma/M_{\mathrm{P}}}\,\sqrt{\tilde{\gamma}}\,(N+Me^{\alpha\sigma/M_{\mathrm{P}}}\mathcal{K})\,,\\
\mathcal{S}_{2} & = & -m^{2}c_{2}e^{2\sigma/M_{\mathrm{P}}}\,\sqrt{\tilde{\gamma}}\,\left[N\mathcal{K}+\frac{1}{2}Me^{\alpha\sigma/M_{\mathrm{P}}}(\mathcal{K}^{2}-\mathcal{K}^{i}{}_{j}\mathcal{K}^{j}{}_{i})\right]\,,\\
\mathcal{S}_{3} & = & -m^{2}c_{3}e^{\sigma/M_{\mathrm{P}}}\sqrt{\gamma}\,(N\,\mathfrak{K}+Me^{\alpha\sigma/M_{\mathrm{P}}})\,,\\
\mathcal{S}_{4} & = & -m^{2}c_{4}\sqrt{\gamma}N\,. \label{eqn:def-S4}
\end{eqnarray}
The contribution from $\mathcal{S}_{4}$ gives rise to a cosmological
constant term. Furthermore, it is clear, as expected, that also in
the metric formalism the graviton mass term in the action, $\sum_{i=0}^{4}\mathcal{S}_{i}$,
is linear in the lapses and does not depend on the shift variables.
This is a consequence of the Lorentz violations in the gravity sector.

\section{Self-accelerating de Sitter cosmology}

\label{sec:desitter}

As a first step, we investigate Friedmann Lema{\^i}tre Robertson Walker (FLRW) backgrounds, and in particular show that de Sitter solutions act as late-time attractors of the theory. We then move on to study linear perturbations about these attractor solutions. We find out that the theory leads to well behaved situations in a wide region of the parameter space.

\subsection{Attractor behavior}

We base our procedure on a flat FLRW ansatz, 
\begin{equation}
N=N(t)\,,\quad N^{i}=0\,,\quad\gamma_{ij}=a^{2}(t)\delta_{ij}\,,\quad\sigma=\sigma(t)\,,\quad\lambda=\lambda(t)\,,\quad\lambda^{i}=0\,, \label{eqn:FLRWansatz}
\end{equation}
for which it is also convenient to introduce the quantities 
\begin{equation}
X\equiv\frac{a_{f}}{a}e^{\sigma/M_{\mathrm{P}}}\,,\quad r\equiv\frac{1}{X}\frac{M}{N}e^{(1+\alpha)\sigma/M_{\mathrm{P}}}\,,
\end{equation}
where $a_{f}=1$ and $M=1$ are the constant scale factor and the
constant lapse function for the fiducial metric. It can be further shown (see appendix \ref{sec:lambda}), that the value $\lambda (t) = 0$ is imposed on any such background. By setting this in the equation of motion for $\lambda$,
we obtain 
\begin{equation}
\frac{1}{N}\frac{d}{dt}\left[a^{4+\alpha}X^{1+\alpha}\,J\right]=0\,,
\end{equation}
where 
\begin{equation}
J\equiv c_{0}X^{3}+3c_{1}X^{2}+3c_{2}X+c_{3}\,.
\end{equation}

If $\alpha\ne4$ then the system thus approaches either $X^{1+\alpha}=0$ or $J=0$. Since $X^{1+\alpha}=0$ would lead to a strong coupling, we have to choose the initial condition of the system within the basin of the attractor at $J=0$ so that the $J$ approaches zero at late time. As a direct consequence we can safely set $dX/dt=0$ on this late-time attractor.

On the other hand, if $\alpha=-4$ then the above equation becomes
\begin{equation}
(c_{1}X^{2}+2c_{2}X+c_{3})\frac{dX}{dt}=0\,,
\end{equation}
and is satisfied by any constant value of $X$. Both cases $\alpha \neq -4$ and $\alpha = -4$ thus admit a late-time de Sitter attractor.

\subsection{de Sitter attractor solution}

For $\alpha\ne-4$, by setting $J=0$, i.e. 
\begin{equation}
c_{0}X^{3}+3c_{1}X^{2}+3c_{2}X+c_{3}=0\,,\label{eqn:J=00003D0_dS}
\end{equation}
and thus $dX/dt=0$, the independent background equations of motion
are 
\begin{eqnarray}
(6-\omega)H^{2} & = & (c_{1}X^{3}+3c_{2}X^{2}+3c_{3}X+c_{4})m^{2}\,,\label{eqn:Friedmann_dS}\\
r-1 & = & \frac{2\omega}{6-\omega}\frac{c_{1}X^{3}+3c_{2}X^{2}+3c_{3}X+c_{4}}{(c_{1}X^{2}+2c_{2}X+c_{3})X}\,,\label{eqn:r-1_dS}
\end{eqnarray}
where $H$ ($>0$) is the Hubble expansion rate. 

For $\alpha=-4$, by setting $dX/dt=0$, the independent background equations of motion
are again (\ref{eqn:Friedmann_dS}) and (\ref{eqn:r-1_dS}) above. 

In analogy with the standard case, we can rewrite equation (\ref{eqn:Friedmann_dS}) as
\begin{equation}
3H^{2}  = 8\pi G_{\textrm{eff}}(\rho_g+\rho_\Lambda)
\end{equation}
with 
\begin{equation}
\rho_g  =  \frac{M_{\mathrm{P}}^2m^{2}}{2} (c_{1}X^{3}+3c_{2}X^{2}+3c_{3}X)\,,\quad \rho_\Lambda = \frac{M_{\mathrm{P}}^2m^{2}}{2}c_4 \,, \quad\textrm{ and } \quad G_\textrm{eff} = \frac{1}{8\pi M_\mathrm{P}}\frac{6}{(6-\omega)}\,.
\label{geff}
\end{equation}
As already mentioned after (\ref{eqn:def-S4}), the contribution from $\mathcal{S}_{4}$ gives rise to a cosmological constant and $\rho_{\Lambda}$ corresponds to the vacuum energy density. The positivity of the effective gravitational constant for the FLRW background thus requires that $\omega<6\,$. In both cases $\alpha\ne -4$ and $\alpha=-4$, if we set $\rho_{\Lambda}=0$, i.e. 
\begin{equation}
c_{4}=0\,,\label{eqn:cc=00003D0}
\end{equation}
then the solution represents a self-accelerating de Sitter universe.

If one takes the limit
$r = 1$,
Minkowski solutions are approached. Indeed, in this limit, Eq.\ (\ref{eqn:r-1_dS}) implies that either 
\begin{equation}
\omega = 0 \quad \textrm{or} \quad (c_{1}X^{3}+3c_{2}X^{2}+3c_{3}X+c_{4}) = 0 \,.
\end{equation} 
While the first option is obviously a case of infinitely strong coupling and can be excluded from our study, the second leads to $H = 0$ and thus to a Minkowski solution through Eq.\ (\ref{eqn:Friedmann_dS}). Furthermore, by considering the ratio of equations (\ref{eqn:Friedmann_dS}) and (\ref{eqn:r-1_dS}), 
\begin{equation}
\frac{H^2}{r-1} = \frac{m^2}{2\omega}\left(c_1 X^2 + 2c_2 X + c_3\right)X\,,\label{eqn:minkowski_ratio}
\end{equation}
we see that the ratio $H^2/(r-1)$ can be kept finite independently of $H^2$ and $r-1$ -- if one excludes the fine-tuned case in which the right hand side vanishes. We will confirm in Sec.\ \ref{sec:minkowski} that the limit $r\to 1$ corresponds indeed to two branches of Minkowski background solutions, and show that, similarly to the de Sitter solutions, they admit stable perturbations.

\subsection{Stability of self-accelerating de Sitter attractor}

\label{sec:stability}

We first define a set of perturbations of the metric, valid both on a de Sitter background and also on more general setups. We decompose the perturbations based on representations of the spatial rotations; the symmetry of the background then ensures that the different modes decouple at linear level. The metric perturbations are thus given by
\begin{eqnarray}
\delta g_{00} &=& -2\,N^2\,\Phi\,,\nonumber\\
\delta g_{0i} &=& N\,a\,\left(\partial_i B+B_i\right)\,,\nonumber\\
\delta g_{ij} &=& a^2 \left[2\,\delta_{ij}\psi +\left(\partial_i\partial_j-\frac{\delta_{ij}}{3}\partial^k\partial_k\right)E+\partial_{(i}E_{j)}+h_{ij}\right]\,,
\label{eqn:perturbations_def}
\end{eqnarray}
where the latin indices are raised by $\delta^{ij}$ and $h_{ij}$, $E_i$, and $B_i$ obey tracelessness and transversality, 
i.e.\ $\delta^{ij}h_{ij} = \partial^ih_{ij} = \partial^i E_i = \partial^i B_i=0$. The quasidilaton scalar field and the Lagrange multipliers are also perturbed as
\begin{equation}
\sigma = \sigma_0 + \delta\sigma\,,\quad
\lambda = \delta\lambda\,,\quad \lambda_i = \partial_i \delta\lambda_L + \delta\lambda_i\,,
\end{equation}
where $\delta\lambda_i$ also obeys a transversality condition, $\partial^i\delta\lambda_i = 0$, and $\sigma_0$ denotes the background value of the field $\sigma$. 

\subsubsection{Case $\alpha\protect\neq-4$\label{subsec:alphaNEQM4}}

On the self-accelerating de Sitter background, with $\alpha\ne-4$,
one can eliminate ($c_{0}$, $c_{1}$, $c_{2}$, $c_{4}$) by using
(\ref{eqn:J=00003D0_dS}), (\ref{eqn:Friedmann_dS}), (\ref{eqn:r-1_dS})
and (\ref{eqn:cc=00003D0}). There are two propagating tensor modes
with the dispersion relation of the form $\omega^{2}=k^{2}/a^{2}+H^{2}\mu_{dS}^{2}$,
and there is no propagating vector mode. There is one propagating
scalar mode with the dispersion relation of the form $\omega^{2}=c_{dS}^{2}k^{2}/a^{2}+H^{2}\nu_{dS}^{2}$.
Here, $c_{dS}^{2}$, $\mu_{dS}^{2}$ and $\nu_{dS}^{2}$ are functions
of ($c_{3}$, $\omega$, $\alpha$, $r$, $X$, $m^{2}/H^{2}$). The
no-ghost condition for the scalar mode is simply $\omega>0$. By choosing
$c_{3}$ as 
\begin{equation}
c_{3}=\frac{(r+\alpha+7)\omega+6(1-r)}{(1-r)X}\frac{H^{2}}{m^{2}}\,,\label{eqn:c3_soundspeed}
\end{equation}
one can set $c_{dS}^{2}=1$. This condition is not 
a fundamental one but makes various expressions simple. In this case the dependence of $\mu_{dS}^{2}$ 
and $\nu_{dS}^{2}$ on $X$ and $m^{2}/H^{2}$ falls off, they thus
depend only on ($\omega$, $\alpha$, $r$), and it is easy to show
that there is a regime of parameters in which $\mu_{dS}^{2}>0$ and
$\nu_{dS}^{2}>0$. In fact, under the condition (\ref{eqn:c3_soundspeed}), we have
\begin{equation}
S_{{\rm sc}}=\frac{1}{2}\int d^{4}x\,Na^{3}\,\omega\left[\left(\frac{1}{N}\frac{\partial(\delta\sigma)}{\partial t}\right)^{2}-\frac{1}{a^{2}}\,[\partial_{i}(\delta\sigma)]^{2}-\nu_{dS}^{2}\,H^{2}\,(\delta\sigma)^{2}\right]\,,
\end{equation}
where we have introduced $\sigma=\sigma(t)+\delta\sigma$, and defined
\begin{eqnarray}
\nu_{dS}^{2} & = & \frac{1}{8[\left(4+\alpha\right)r-\alpha-\tfrac{1}{2}\omega-4]^{2}\left(r-1\right)}\,\{24\,\alpha^{3}+\left(8\,\omega^{2}-12\,\omega+264\right)\alpha^{2}+\left(86\,\omega^{2}-132\,\omega+888\right)\alpha\nonumber \\
 &  & {}-3\,\omega^{3}+230\,\omega^{2}-252\,\omega+936\nonumber \\
 &  & {}+[3\,\omega^{3}-354\,\omega^{2}+156\,\omega-1608-72\,\alpha^{3}-\left(16\,\omega^{2}+672\right)\alpha^{2}+\left(60\,\omega-152\,\omega^{2}-1920\right)\alpha]\,r\nonumber \\
 &  & {}+72\,\bigl[\alpha^{2}+\left(\tfrac{5}{36}\,\omega^{2}+\tfrac{1}{6}\omega+\tfrac{14}{3}\right)\alpha+\tfrac{7}{12}\,\omega^{2}+\tfrac{1}{3}\,\omega+\tfrac{11}{3}\bigr]\left(4+\alpha\right)r^{2}-24\,\left(\tfrac{1}{12}\omega^{2}+\alpha+1\right)\left(4+\alpha\right)^{2}r^{3}\}\,.
\end{eqnarray}

In the same case, 
i.e.\ with (\ref{eqn:c3_soundspeed}) and thus $c_{dS}^{2}=1$, the action for the tensor modes reduces to
\begin{equation}
S_{{\rm ten}}=\frac{\Mpl^{2}}{4}\sum_{\lambda={+},{\times}}\int d^{4}x\,Na{}^{3}\left[\left(\frac{1}{N}\frac{\partial h_{\lambda}}{\partial t}\right)^{2}-\frac{1}{a^{2}}\,(\partial_{i}h_{\lambda})^{2}-\mu_{dS}^{2}\,H^{2}\,h_{\lambda}^{2}\right],
\end{equation}
where
\begin{equation}
\mu_{dS}^{2}=\frac{[\alpha+6-\left(\alpha+4\right)r]\,\omega}{2\,(r-1)}\,.
\end{equation}
There is a large region of parameters that ensure stability. For instance, we find that $\mu_{dS}^{2}$ and $\nu_{dS}^{2}$ are positive for 
$0<\omega\ll 1$, $0<r\ll 1$, $\alpha<-6$. 

In the limit $r \to 1$, both $\mu_{dS}^2 H^2$ and $\nu_{dS}^2 H^2$ remain finite, as they are proportional to the ratio $H^2/(r-1)$ -- see Eq.\ (\ref{eqn:minkowski_ratio}). Moreover, one can see that the condition ensuring $c_{dS}^2 = 1$, Eq.\ (\ref{eqn:c3_soundspeed}), also remains finite.

\subsubsection{Case $\alpha=-4$\label{subsec:alphaEQM4}}

On the self-accelerating de Sitter background with $\alpha=-4$, one
can eliminate ($c_{1}$, $c_{2}$, $c_{4}$) by using (\ref{eqn:Friedmann_dS}), (\ref{eqn:r-1_dS}) and (\ref{eqn:cc=00003D0}).
There are two propagating tensor modes with the dispersion relation
of the form $\omega^{2}=k^{2}/a^{2}+H^{2}\mu_{dS}^{2}$, and there
is no propagating vector mode. There is one propagating scalar mode
with the dispersion relation of the form $\omega^{2}=c_{dS}^{2}k^{2}/a^{2}+H^{2}\nu_{dS}^{2}$.
Here, $c_{dS}^{2}$, $\mu_{dS}^{2}$ and $\nu_{dS}^{2}$ are 
\begin{equation}
c_{dS}^{2}=\frac{(\omega+10)\omega}{(6-\omega)^{2}}\,,\quad\nu_{dS}^{2}=-\frac{3(\omega+6)\omega}{2(6-\omega)}\,,\quad\mu_{dS}^{2}=\frac{(\omega r^{2}+2\omega r-6r^{2}-\omega+12r-6)}{2(r-1)}+\frac{c_{3}}{2}\frac{m^{2}}{H^{2}}X(r-1)\,.
\end{equation}
The no-ghost condition for the scalar mode is simply $\omega>0$.
It is easy to see that $\nu_{dS}^{2}<0$ as far as $0<\omega<6$.

To summarize, one can easily find a regime of parameters with ($\alpha\ne-4$) in which the self-accelerating de Sitter solution is stable, while the infinitely fine-tuned case ($\alpha=-4$) is unstable in the IR. However, as the time scale of this instability is of order of the age of the Universe, it may not be problematic. In the Minkowski limit, $H^2 \nu_{dS}^2 $ vanishes, thus in this limit the solution can be considered as safe. Indeed, we note that also $H^2 \mu_{dS}^2$ remains finite in the $r \to 1$ limit, showing that the Minkowski limit is smooth for any value of $\alpha$.

\subsection{Summary of consistency conditions}

The first constraints we set are $r>0$ and $0<\omega<6$. While the former is a consequence of the positivity of the lapse function and the scale factor, the latter stems from considering the effective gravitational constant on the late-time de Sitter attractor and requiring its positivity [Eq.\ (\ref{geff})], together with one of the no-ghost conditions. We further have required that there is altogether no gradient or tachyonic instability for the perturbation modes.

As a supplementary condition we have set the sound speed in the scalar sector to $c_{dS}^2 = 1$. While this constraint is not fundamental, it allowed us to simplify the calculations in a relevant way. We do not lose generality by doing so, in the sense that, even under this constraint, the allowed parameter region is still large. By lifting the aforementioned constraint we can thus only expect that an even wider parameter range is allowed. 

We further explore different cases under the $c_{dS}^2 = 1$ assumption. One can split the parameter space in two different regions depending on the value of $\alpha$. If $\alpha<-6$, both the regions $0<r<1$ and $r>1$ become available. In particular, as seen previously, it allows for the particular region $0<\omega\ll1$, $0<r\ll 1$. For $-6\leq \alpha$, the region $0<r<1$ becomes unavailable, but $r>1$ preserves a wide stable region. If one wants to consider the special case $\alpha = -4$, one must do so earlier in the analysis, as one of the equations of motion [Eq. (\ref{eqn:J=00003D0_dS})] is removed. 

As one can see by taking any $\alpha < -4$, both $\mu_{dS}^2$ and $\nu_{dS}^2$ remain positive for any value of $r\gg1$, thus there is no general upper limit for $r$. Similarly, as shown by the limit $\omega\to 6$, there is, for any value of $\alpha \gg 1$, always a range of $r>1$ that allows for all conditions to be satisfied. Thus, there is also no general upper limit for $\alpha$.

Finally, we have seen that the limit $r \to 1$ corresponds to a Minkowski solution. In particular, we have shown that the quantities $\mu_{dS}^2H^2$ and $\nu_{dS}^2H^2$ remain finite in this limit, thus providing a smooth way from de Sitter to Minkowski.

\begin{figure}[h]
 \centering
  \includegraphics[width=6.5cm]{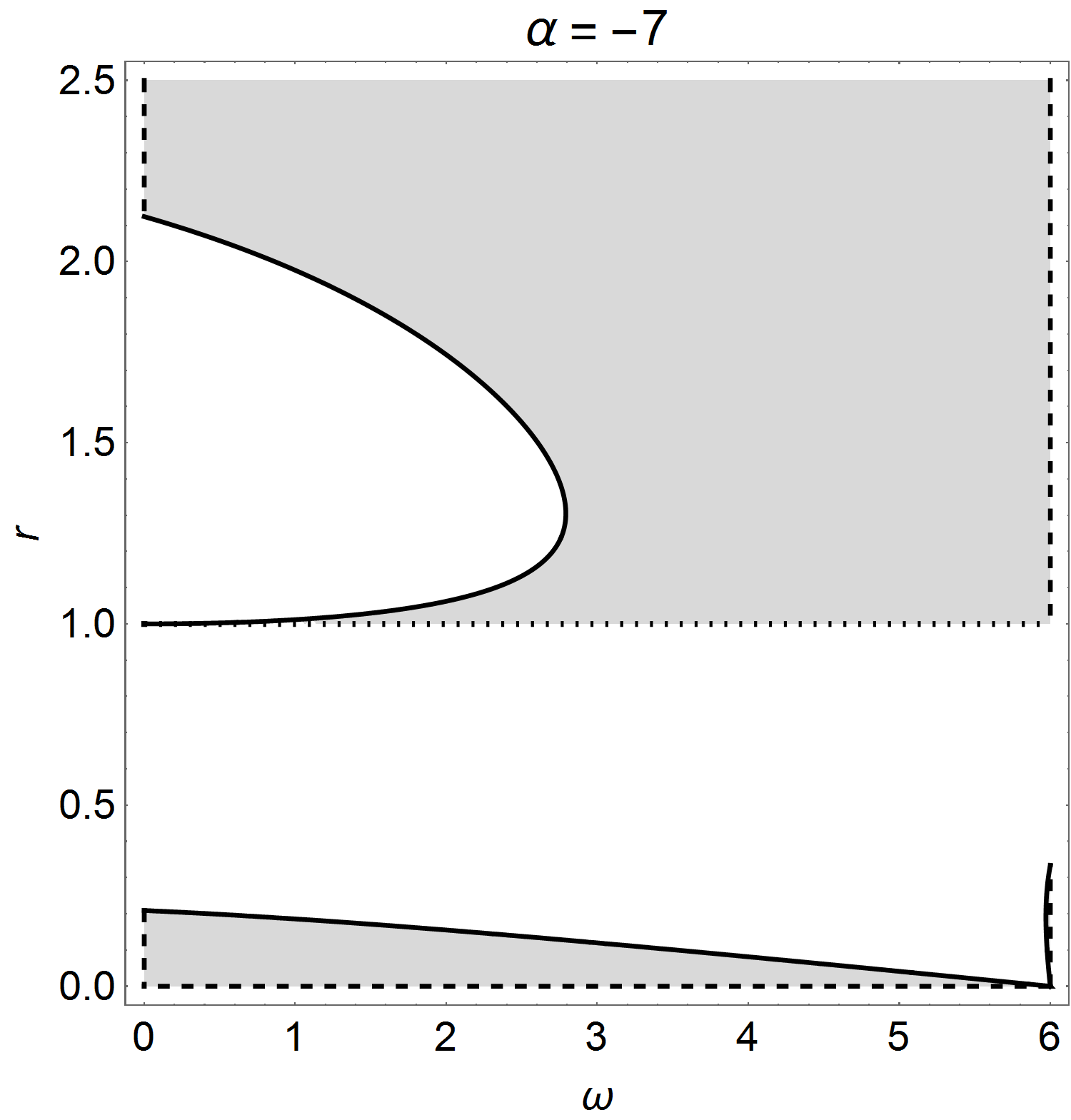}
  \includegraphics[width=6.5cm]{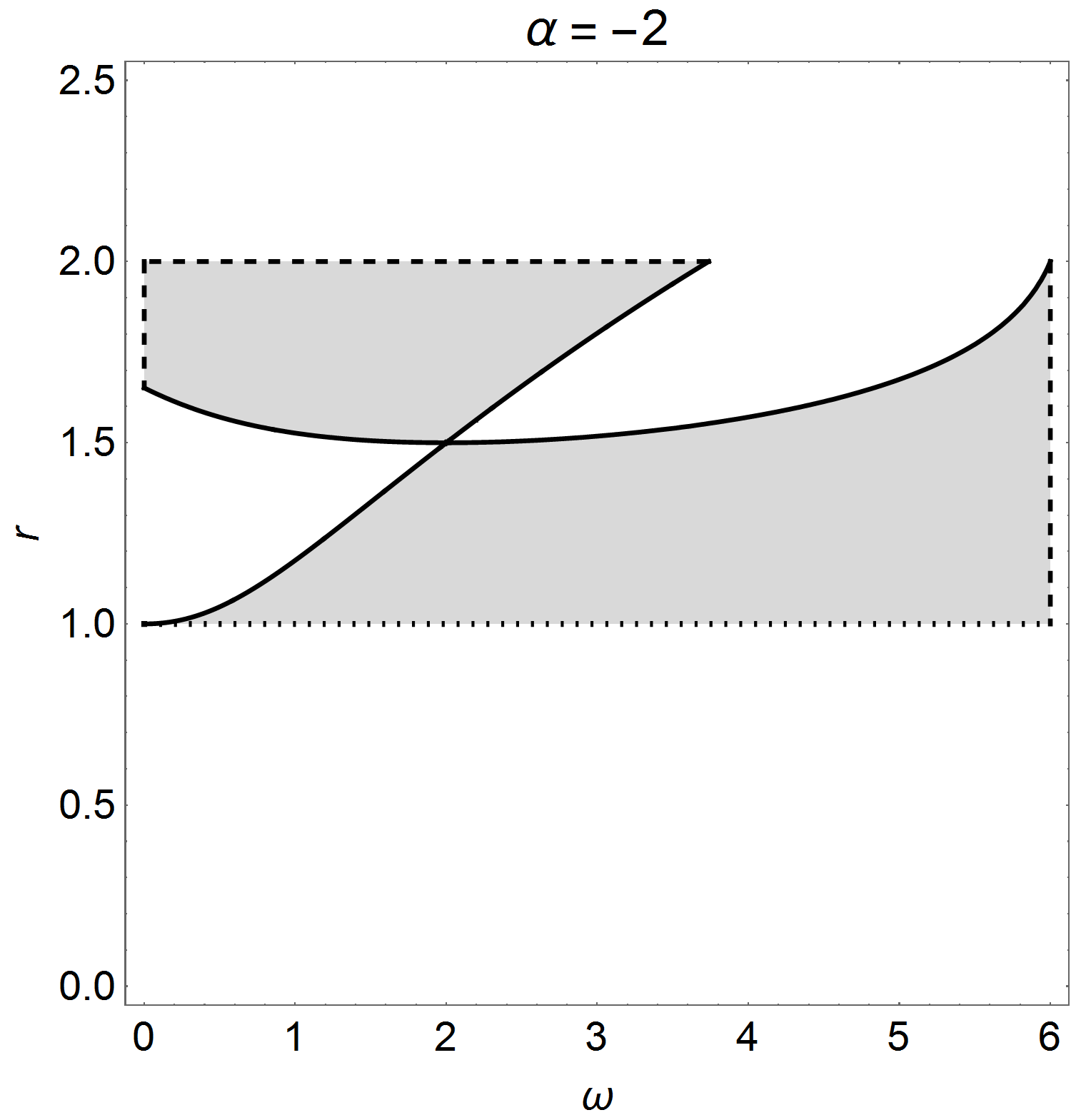}\\
  \vspace{0.4cm}
  \includegraphics[width=6.5cm]{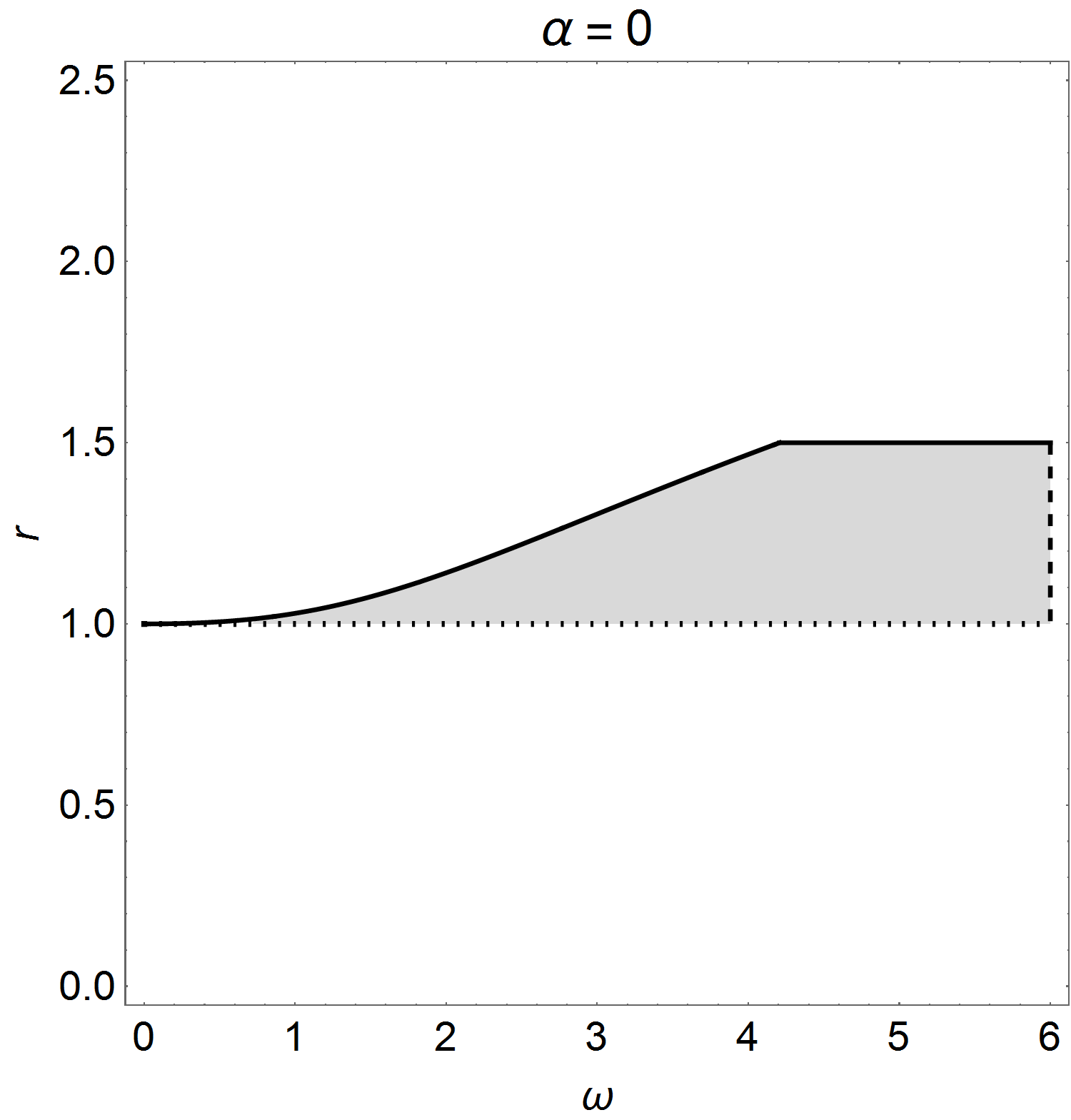}
  \includegraphics[width=6.5cm]{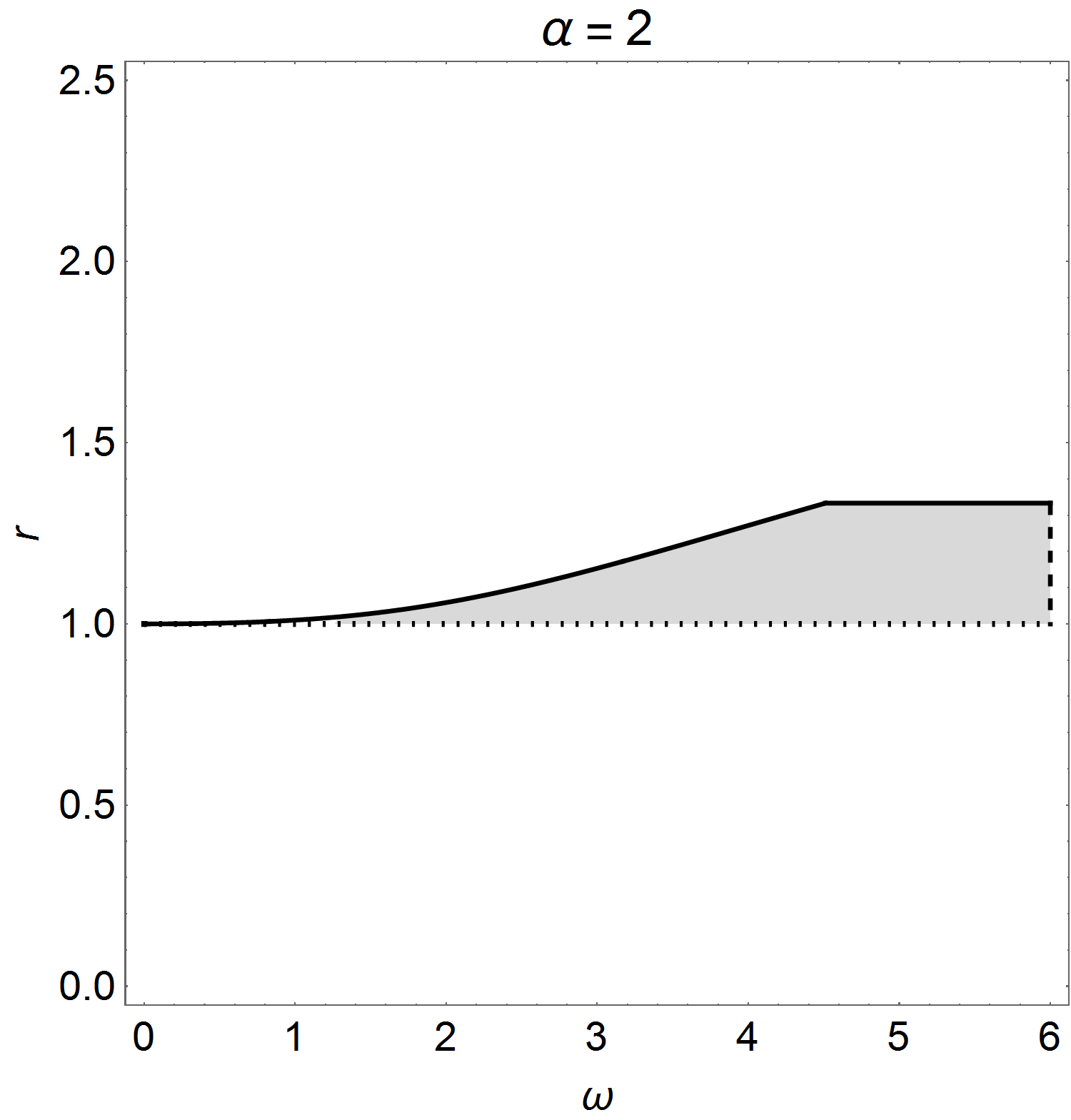}
 \caption{Four examples of allowed regions in the parameter space of the theory, for the values $\alpha = -7$, $\alpha = -2$, $\alpha=0$, and $\alpha= 2$. The shaded areas denote the regions in the $(\omega,r)$ plane for which both $\nu_{dS}^2 >0$ and $\mu_{dS}^2 >0$, under the condition $c_{dS}^2 = 1$ and $c_4=0$. Lifting one or both of these two conditions would only enlarge the allowed region. Full lines correspond to either one of the masses becoming zero and are thus included in the allowed region. Dashed lines are not permitted. In particular, the parameter space has the following restrictions: $0<\omega<6$ due to a no-ghost condition and positivity of the effective gravitational constant, $r>0$ from the positivity of the lapse function and the scale factor. The dotted line at $r = 1$ corresponds to the Minkowski limit.}\label{allowedparameterspace}
\end{figure}

We illustrate these results by depicting in Fig. \ref{allowedparameterspace} the allowed parameter space for four different values of $\alpha$, under the constraint $c_{dS}^2=1$ and $c_4=0$. 

\section{Minkowski limit}

\label{sec:minkowski}

As a second step in the study of background solutions and their stability we choose to investigate Minkowski solutions and compare them with the Minkowski limit of the de Sitter solutions investigated in the previous section. From the background equations of motion we obtain four distinct branches of Minkowski solutions. Two of these branches correspond to the safe limit obtained from the de Sitter solution, while the other two branches require an infinite fine-tuning and are disconnected from the de Sitter solutions.

\subsection{Background}

We use the FLRW ansatz (\ref{eqn:FLRWansatz}) already exposed in previous section. Minkowski solutions are then specified by 
\begin{equation} 
a = a_M = const.\,,\quad N = N_M = const.\,,\quad \sigma_0 = \sigma_{0M} = const.\,, 
\end{equation}
and we adopt the subscript $~_M$ to indicate their respective constant values. 
The 4-metric can thus be written
\begin{equation}
ds_M^2 =  -N_M^2 dt^2 + a_M^2 \delta_{ij} dx^i dx^j\,.
\end{equation}

Under such an ansatz, the Friedmann equation leads to 
\begin{equation}
c_{1}X_{M}^{3}+3c_{2}X_{M}^{2}+3c_{3}X_{M}+c_{4}=0\,,
\end{equation}
where  $X_{M}$ is the constant value of $X$ on the Minkowski solution, while the second Einstein equation (often called a dynamical equation) and the equation of motion for the quasidilaton lead to 
\begin{equation}
(r_{M}-1)(c_{1}X_{M}^{2}+2c_{2}X_{M}+c_{3})=0\,,\quad(\alpha+4)(c_{0}X_{M}^{3}+3c_{1}X_{M}^{2}+3c_{2}X_{M}+c_{3})=0\,,
\end{equation}
where $r_{M}$ is the value of $r$ on the Minkowski solution. In principle the Minkowski solution thus has four branches: 
\begin{itemize}
\item[(ia)] $c_{1}X_{M}^{3}+3c_{2}X_{M}^{2}+3c_{3}X_{M}+c_{4}=r_{M}-1=c_{0}X_{M}^{3}+3c_{1}X_{M}^{2}+3c_{2}X_{M}+c_{3}=0$. 
\item[(ib)] $c_{1}X_{M}^{3}+3c_{2}X_{M}^{2}+3c_{3}X_{M}+c_{4}=r_{M}-1=\alpha+4=0$. 
\item[(iia)] $c_{1}X_{M}^{3}+3c_{2}X_{M}^{2}+3c_{3}X_{M}+c_{4}=c_{1}X_{M}^{2}+2c_{2}X_{M}+c_{3}=c_{0}X_{M}^{3}+3c_{1}X_{M}^{2}+3c_{2}X_{M}+c_{3}=0$. 
\item[(iib)] $c_{1}X_{M}^{3}+3c_{2}X_{M}^{2}+3c_{3}X_{M}+c_{4}=c_{1}X_{M}^{2}+2c_{2}X_{M}+c_{3}=\alpha+4=0$. 
\end{itemize}

As we have seen in the previous section, the two first branches, (ia) and (ib), correspond to the generic limit $r \to 1$ of the de Sitter solutions, while the latter branches, (iia) and (iib), require a supplementary infinite fine-tuning -- from the point of view of the de Sitter solutions -- to set the combination $(c_{1}X_{M}^{2}+2c_{2}X_{M}+c_{3})$ to zero. Moreover, the branches (iia) and (iib) with $r_M\ne 1$ are disconnected from the de Sitter solutions. We will see next that these two fine-tuned branches (iia) and (iib) do not propagate any scalar mode and thus exhibit infinite strong coupling, while the first two cases allow for a stable scalar perturbation mode. 

\subsection{Perturbations}

We parametrize the perturbations as in the de Sitter case [Eq.\ \ref{eqn:perturbations_def}], adding subscripts $~_M$ to the quantities $a$, $N$, and $\sigma_0$ as done for the study of the background. Next, we replace all four branches of the Minkowski solution in the perturbed action, and then proceed to integrate out any nondynamical degree of freedom. We first find that, in all four branches, there are two propagating tensor modes with the dispersion relation of the form $\omega^{2}=k^{2}+m^{2}\mu_{M}^{2}$, and there is no propagating vector mode. 

The scalar modes require more scrutiny. Generically, the equations stemming from the variation of $\delta\lambda$, $\Phi$, and $B$ can be solved and end up fixing the respective value of these three perturbations. In contrast to this, $\delta\lambda_L$ always appears as a Lagrange multiplier. The equation stemming from its variation thus imposes a constraint on the remaining variables, and allows us to fix one of them. For the cases (iia) and (iib), there subsists no propagating scalar mode. On the other hand, for the cases (ia) and (ib), there is one propagating scalar mode with the dispersion relation of the form $\omega^{2}=c_{M}^{2}k^{2}+m^{2}\nu_{M}^{2}$. The no-ghost condition for the scalar mode for the cases (ia) and (ib) is simply $\omega>0$. 

We already showed that the branches (ia) and (ib) are the ones corresponding to the smooth Minkowski limit of the de Sitter attractor solutions. As a consistency check, one can in fact compare the quadratic action for perturbations around the Minkowski background, taking $c_4 = 0$, to the action obtained in the Minkowski limit of the de Sitter background. For $\alpha\ne -4$, we find it convenient to eliminate $H^2$ in favor of $c_3m^2$ by using (\ref{eqn:c3_soundspeed}) before taking the limit $r\to 1$. For $\alpha \neq -4$, one then obtains 
\begin{equation}
 \lim_{r\to 1}H^2\nu_{dS}^2 = m^2\nu_{M}^2 = -\frac{6 c_3 m^2 X_M}{\omega(8+\alpha)}\,,\quad \lim_{r\to 1}H^2\mu_{dS}^2 = m^2\mu_{M}^2 = -\frac{c_3 m^2 X_M}{8+\alpha}\,,\quad (\mbox{for}\ \alpha\ne -4\,,\  c_{\rm dS}^2=c_{M}^2=1\,,\  c_4=0)\,.
\end{equation}
For $\alpha = -4$, one obtains
\begin{equation}
 \lim_{r\to 1}H^2\nu_{dS}^2 = m^2\nu_{M}^2 = 0\,,\quad \lim_{r\to 1}H^2\mu_{dS}^2 = m^2\mu_{M}^2 = \frac{m^2}{6}(c_1X_M^3-3c_3X_M)\,,\quad (\mbox{for}\  \alpha= -4\,,\  c_4=0)\,.\label{eqn:min_limit}
\end{equation}
Therefore the Minkowski limit is well defined and smooth.

In conclusion, while the scalar sector is infinitely strongly coupled for the cases (iia) and (iib) that are disconnected from the de Sitter solutions, the cases (ia) and (ib) are stable and correspond to the smooth Minkowski limit of the de Sitter case.

\section{Summary and discussion}
\label{sec:summary}

We have proposed a new theory of massive gravity which possesses three propagating degrees of freedom at fully nonlinear level. The theory can be considered as a quasidilaton extension \cite{DAmico:2012hia} of the MTMG that was recently proposed \cite{DeFelice:2015hla, DeFelice:2015moy}. In \cite{DeFelice:2015hla, DeFelice:2015moy} the MTMG with 2 degrees of freedom was obtained in the so-called unitary gauge by introducing a preferred frame and adding two additional constraints to the Hamiltonian of a precursor theory that originally has three degrees of freedom. In the present paper, we first covariantized the precursor theory with a Minkowski fiducial metric by introducing St\"{u}ckelberg fields. We then introduced a quasidilaton scalar field that respects a global symmetry mixing the quasidilaton and St\"{u}ckelberg fields. From the quasidilaton extension of the precursor theory constructed in this way, we eliminated one propagating degree of freedom by adding two additional constraints that were carefully chosen. We then ended up with a theory with 3 degrees of freedom including the quasidilaton. We call it the \textit{minimal quasidilaton}. As in the MTMG, the necessary Lorentz violation is limited to the gravity sector, thus appearing only at length scales of order $1/m$ or larger, i.e.\ at cosmological scales. 

After constructing the theory of minimal quasidilaton, we found an attractor solution that represents a self-accelerating de Sitter universe and that is stable in a range of parameters. Subsequently, we investigated the Minkowski limit of the de Sitter solutions and showed that the limit is smooth.  Alternative and disconnected Minkowski branches are shown to be possible, however via fine-tuning and at the price of infinite strong coupling. Therefore, the only consistent Minkowski solutions are those that are smoothly connected to de Sitter solutions. 

The theory was constructed in the unitary gauge but it is straightforward to covariantize the action by introducing St\"{u}ckelberg fields as we have done for the precursor theory. The covariant action is expected to be useful, 
e.g.\ for the analysis of the decoupling limit. 

One of the important phenomenological questions in modified gravity theories is how screening mechanisms work. In this respect, it is interesting to include shift-symmetric Horndeski terms for the quasidilaton scalar field to the system \cite{DeFelice:2013dua}. This should suffice to screen the fifth force due to the quasidilaton scalar field. Since the minimal quasidilaton does not contain an extra degree of freedom that stems from the massive gravity part, the addition of shift-symmetric Horndeski terms for the quasidilaton scalar is expected to be sufficient for the recovery of general relativity within the Vainshtein radius. It is worthwhile to investigate this issue in detail.

Inclusion of matter in the cosmological context is also an important issue since the evolution of cosmological perturbations is expected to be different from general relativity. In the MTMG without the quasidilaton, it is known that there are two distinct branches of cosmological solutions. In the so-called self-accelerating branch of the MTMG, the evolution of scalar- and vector-type linear perturbations as well as the FLRW background is exactly the same as that in $\Lambda$ cold dark matter ($\Lambda$CDM). In the other branch of the MTMG, on the other hand, while the vector-type linear perturbations are absent as in the $\Lambda$CDM, the scalar-type linear perturbations tend to show a weaker gravity at late time in accord with some of recent observations \cite{DeFelice:2016ufg}. It would be interesting to study how the scalar-type cosmological perturbations behave in the minimal quasidilaton with matter.

While the recent direct detection of gravitational waves by aLIGO put an upper bound on the mass of gravitational waves as $m_{\textrm{gw}} < 1.2 \times 10^{-22}\,eV$ \cite{bib:LIGO}, in the context of the self-accelerating solution of the MTMG and the minimal quasidilaton the current acceleration of the cosmic expansion suggests even lower value of order $H_0 \sim 10^{-33}\, eV$ . Therefore it is certainly important to push forward the observational upper bound as much as we can. From the theoretical point of view, it is also important to investigate whether a small graviton mass is technically natural in the context of the MTMG and the minimal quasidilaton, as in the dRGT theory \cite{deRham:2012ew,deRham:2013qqa}.

\acknowledgments
A.D.F.\ was supported by JSPS KAKENHI Grant No.\
16K05348, No.\ 16H01099. The work of S.M.\ was supported by Japan Society for
the Promotion of Science (JSPS) Grants-in-Aid for Scientific Research
(KAKENHI) No.\ 24540256, No.\ 17H02890, No.\ 17H06359, No.\ 17H06357, and by World Premier International Research Center Initiative (WPI), MEXT, Japan.

\appendix

\section{Value of $\lambda$ on the background}

\label{sec:lambda}

For the general FLRW ansatz (\ref{eqn:FLRWansatz}), we show in this appendix that combining equations of motion and time derivatives of some of them, we can obtain an algebraic equation for $\lambda$ of the form 
\begin{equation}
E_B =A\lambda=0\,,
\end{equation}
where $A$ is a linear polynomial of $\lambda$ whose coefficients depend on $H$, $X$, $dX/dt$ but do not depend on $dH/dt$, $d^{2}X/dt^{2}$, $d^{2}r/dt^{2}$ etc. This
implies $\lambda=0$ on any FLRW-type background unless we set $A=0$, which would lead to inconsistencies. The combination of equations is chosen as
\begin{equation}
E_B = \dot{E}_0 + b_1 E_0 + b_2 E_1 + b_3 E_\sigma + b_4 \dot{E}_\lambda + b_5 E_\lambda\,,
\end{equation}
where $E_0$ is the Friedmann equation, $E_1$ the second Einstein equation, $E_\sigma$ the equation for the quasidilaton field, $E_\lambda$ the equation for $\lambda$, and dots denote time derivatives. We use the freedom in the $b_i$ ($i = 1,\,\cdots,5$) to eliminate successively the dependence in $\ddot{\sigma}$, $\dot{H}$, etc., until only the desired dependence remains. In fact, after having solved for $b_3$ in order to eliminate the dependence in $\ddot{\sigma}$, one can greatly simplify the intermediate steps by taking the special value
\begin{equation}
b_5 = \frac{m^2 r x \lambda}{\omega}\,.
\end{equation}
By doing so, if one moves on to suppress $\dot{H}$, the simple solution
\begin{equation}
b_2 = -3 H N\,
\end{equation}
is obtained. The resulting expression is devoid of dependence in $\dot{\lambda}$, and is simply a quadratic polynomial in $\lambda$. As a consequence, one can directly use the two remaining variables $b_1$ and $b_4$ to eliminate the constant term and the quadratic term.

\end{document}